\pdfoutput=1
\documentclass[a4paper,10pt]{article}
\usepackage[utf8]{inputenc}
\usepackage{amsthm}
\usepackage{amsfonts}
\usepackage{amssymb}	
\usepackage{amsmath}
\allowdisplaybreaks
\usepackage{mathtools}
\usepackage[english]{babel}
\usepackage{color}
\usepackage{cite}
\usepackage[numbers]{natbib}
\usepackage{slashed}
\usepackage{enumerate}
\usepackage{graphicx}
\usepackage{systeme}
\usepackage{dsfont}
\usepackage{relsize}
\usepackage[multiple]{footmisc}
\usepackage[margin=1in]{geometry}
\usepackage{float}
\usepackage{subcaption}
\usepackage{tikz}
\usepackage[labelformat=simple]{subcaption}

\usepackage{graphicx}
\usepackage{subcaption}
\usepackage{bmpsize}
\usepackage{epstopdf}

\usepackage{bbm}
\usepackage{xcolor,etoolbox}
\usepackage{titling}
\usepackage{authblk}
\newcommand*\samethanks[1][\value{footnote}]{\footnotemark[#1]}

\usepackage{blindtext,graphicx}
\usepackage[absolute]{textpos}
\usepackage{physics}
\usepackage{stmaryrd}

\title{Exploring metamaterials' structures through the relaxed micromorphic model: switching an acoustic screen into an acoustic absorber}
\author{
Gianluca Rizzi\thanks{GEOMAS, INSA-Lyon, Université de Lyon, 20 avenue Albert Einstein, 69621, Villeurbanne Cedex, France}
,\quad
Manuel Collet\thanks{École Centrale de Lyon, 36 avenue Guy de Collongue, Écully 69134, France}
,\quad
Félix Demore\samethanks[2]
,\quad
Bernhard Eidel\thanks{Heisenberg-Group, Institute of Mechanics, Department Mechanical Engineering, University Siegen, \\ \indent Paul-Bonatz-Str. 9-11, 57068, Siegen, Germany}
,\\
Patrizio Neff\thanks{Head of Chair for Nonlinear Analysis and Modelling, Fakultät für Mathematik, Universität Duisburg-Essen, \\ \indent Thea-Leymann-Straße 9, 45127 Essen, Germany}
,\quad
Angela Madeo\samethanks[1]
}

\thanksmarkseries{arabic}
\date{\today}

\makeindex

\begin{document}
\maketitle
\begin{abstract}
\begin{small}
While the design of always new metamaterials with exotic static and dynamic properties is attracting deep attention in the last decades, little effort is made to explore their interactions with other materials. This prevents the conception of (meta-)structures that can enhance metamaterials'~unorthodox behaviours and that can be employed in real engineering applications. In this paper, we give a first answer to this challenging problem by showing that the relaxed micromorphic model with zero static characteristic length can be usefully applied to describe the refractive properties of simple meta-structures for extended frequency ranges and for any direction of propagation of the incident wave. Thanks to the simplified model's structure, we are able to efficiently explore different configurations and to show that a given meta-structure can drastically change its overall refractive behaviour when varying the elastic properties of specific meta-structural elements. In some cases, changing the stiffness of a homogeneous material which is in contact with a metamaterial's slab, reverses the structure's refractive behaviour by switching it from an acoustic screen (total reflection) into an acoustic absorber (total transmission).
The present paper clearly indicates that, while the study and enhancement of the intrinsic metamaterials' properties is certainly of great importance, it is even more challenging to enable the conception of meta-structures that can eventually boost the use of metamaterials in real-case applications.
\end{small}
\end{abstract}

\vspace{2mm}
\section{Introduction}\label{sec:intro}
The last decade has seen the birth of a true research outburst on so-called mechanical metamaterials which are able to show exotic mechanical properties both in the static and dynamic regime.
Theoretical, experimental and numerical studies have flourished all around the world providing new insights in the domain of material properties manipulation which, only few years ago, was thought far from being prone to possible ground-breaking evolutions.
We are today assisting to the conception and subsequent realization of new materials which, simply thanks to their internal architecture, go beyond the materials’~mechanical properties that we are used to know and which, for this reason, are called mechanical metamaterials.
Already in the late 1980s, it was proven that some foams with special internal architecture can give rise to ‘negative Poisson’ effects, i.e., they fatten when stretched, contrarily to what happens to the great majority of known materials which experience a reduction in the cross-section when submitted to tensile loads \cite{lakes1987foam}.
More recently, the frontiers of metamaterials’ conception are rapidly moving forward, giving rise to the manufacturing of always new metamaterials with more and more impressive properties.
It is thus possible today to see 3D-printed pyramids connected by hinges giving rise to a block that is hard like a brick on one side but soft like a sponge on the other \cite{bilal2017intrinsically}, ``unfeelability'' cloaks hiding to the touch objects put below them \cite{milton1995elasticity,kadic2014pentamode}, plastic cubes made out of smaller plastic cubes giving rise to bizarre deformations when squeezed \cite{coulais2016combinatorial}, or even metamaterials exploiting microstructural instabilities to change their mechanical response depending on the level of externally applied load \cite{kochmann2017exploiting}.
When considering the dynamical behaviour of mechanical metamaterials, things become, if possible, even more impressive, given the unbelievable responses that such metamaterials can provide when coming in contact with elastic waves \cite{deymier2013acoustic,hussein2014dynamics,barchiesi2019mechanical}.
It is today possible to find researchers working on metamaterials exhibiting band-gaps \cite{celli2019bandgap,bilal2018architected,liu2000locally,wang2014harnessing,el2018discrete,koutsianitis2019conventional,goh2019inverse,zhu2015study}, cloaking \cite{buckmann2015mechanical,misseroni2016cymatics,norris2014active}, focusing \cite{cummer2016controlling,guenneau2007acoustic}, channelling \cite{kaina2017slow,tallarico2017edge,bordiga2019prestress}, negative refraction \cite{willis2016negative,bordiga2019prestress,zhu2015study}, etc., as soon as they interact with mechanical waves.

Notwithstanding the massive research efforts deployed to unveil new metamaterials’ performances, researchers have just begun to understand the underlying mechanisms, so that “many designs so far have relied on luck and intuition” \cite{bertoldi2017flexible}.

In order to provide deeper theoretical insight into the mechanisms which allow to tailor metamaterials’~mechanical properties, so-called homogenization techniques have been developed which provide rigorous predictions of the macroscopic metamaterials’ mechanical behaviour, when knowing the properties of the base materials and their spatial distribution.
Homogenization techniques have proven their effectiveness for the description of metamaterials’ bulk behaviour in the static and quasi-static regime \cite{bensoussan2011asymptotic,sanchez1980non,allaire1992homogenization,milton2002theory,hashin1963variational,willis1977bounds,pideri1997second,bouchitte2002homogenization,camar2003determination,suquet1985elements,miehe1999computational,geers2010multi,hill1963elastic} as well as, more recently, in the dynamic regime \cite{bacigalupo2014second,chen2001dispersive,boutin2014large,craster2010high,andrianov2008higher,hu2017nonlocal,willis2009exact,willis2011effective,willis2012construction,srivastava2014limit,sridhar2018general}.

While homogenization methods are effective to describe bulk metamaterials' behaviours, they are intrinsically unsuitable to deal with metamaterials of finite size, because the ‘average operations’, on which they are built, make strong use of projection functions (e.g., Bloch-Floquet ones) that are defined in unbounded function spaces \cite{srivastava2017evanescent,sridhar2016homogenization}.

As a result of this gap, the response of finite-size metamaterials’ structures is today mostly explored via direct Finite Element (FEM) simulations that implement all the details of the involved microstructures (e.g., \cite{krushynska2017coupling}).
Despite the precise propagation patterns that these direct numerical simulations can provide, they suffer from unsustainable computational costs.
Therefore, it is impossible today to explore large-scale meta-structures combining metamaterials and classical-materials bricks of different type, size and shape.

A first answer to this problem has been given by the introduction of the so-called relaxed micromorphic model without curvature contribution that has recently proven its effectiveness for the description of the mechanical behaviour of a specific finite-size band-gap metamaterial with tetragonal symmetry \cite{d2019effective}.
In the present paper, we want to move beyond these first encouraging results and show how the relaxed micromorphic model without curvature contribution can be used to characterize many other tetragonal band-gap metamaterials that can be used for acoustic applications.
To this aim, we will apply the inverse fitting procedure presented in \cite{neff2019identification, d2019effective} to different metamaterials thus providing their mechanical description via the relaxed micromorphic model without curvature contribution.
We will then explore how the behaviour of these metamaterials can be profoundly changed by simply coupling them to classical homogeneous materials, realizing what we will call meta-structures, which can have very different mechanical properties when compared to those of the original metamaterial.
The exploration of these new meta-structures is made possible thanks to the simplified structure of the relaxed micromorphic model without curvature contribution that allows quick computations of different structures obtained by embedding the selected metamaterial in different homogeneous materials.
We show how the simple fact of changing the elastic properties of the external homogeneous material allows us to switch the structure's behaviour from total reflection to total transmission and vice-versa.
It is clear that this new possibility of exploring different combinations of finite-size metamaterials and classical-materials opens concrete perspectives for the true employment of metamaterials in engineering design.
Indeed, metamaterials' reflection and transmission properties have been analysed so far, always referring to the particular arrangement of their internal architecture so as to modify the metamaterial's reflection/transmission behavior when the metamaterial is embedded, e.g., in air.
Although such approaches can lead to the design of tunable interfaces allowing total transmission or total reflection depending on the topological microstructure's properties \cite{zhou2007complete,zhao2018tunable,park2019zero}, they are not suitable to explore the behaviour of the same metamaterials when they are used as building blocks of more complex structures that also contain other metamaterials and/or classical-materials elements, especially when these structures are widely extended in space.

In this paper we show how the simple fact of assembling a finite-size metamaterial together with blocks of classical-materials can dramatically change the response of the metamaterial itself in such a way that the same structure may act as a complete absorber (total transmission) or a complete screen (total reflection).
This switch of the structure's dynamical properties is obtained by keeping the same geometry and the same metamaterial, while changing the homogeneous material in which the metamaterial itself is embedded.
These results have been made possible thanks to the use of the relaxed micromorphic model without curvature contribution that, drastically reducing the computational time of the associated numerical simulations, could open the effective exploration of these new meta-structures.
We clearly show that, while the interest of studying the intrinsic metamaterial's properties by engineering its microstructure is certainly of great importance, it is even more important to unveil the effects of its interactions with other finite-size metamaterials and classical-materials bricks.
It is indeed based on these interactions that it is possible to unfold new meta-structures which can further enhance the properties of the base metamaterials thus opening the way to realistic applications.
\subsection{Notation}
\label{sec:notation}
We recall here the notation that we will use throughout the paper.
Let $\mathbb{R}^{3\times 3}$ be the set of all real $3\times 3$ second order tensors which we denote by capital letters.
A simple and a double contraction between tensors of any suitable order is denoted by $\cdot$ and $:$ respectively, while the scalar product of tensors of suitable order is denoted by $\left\langle \cdot, \cdot \right\rangle$.
\footnote{
	For example, $(A\cdot v)_i=A_{ij}v_j$,  $(A\cdot B)_{ik}=A_{ij}B_{jk}$, $A:B=A_{ij}B_{ji}$, $(C\cdot B)_{ijk}=C_{ijp}B_{pk}$, $(C:B)_{i}=C_{ijp}B_{pj}$, $\left\langle v,w\right\rangle = v\cdot w = v_i w_i$, $\left\langle A, B \right\rangle = A_{ij}B_{ij}$, etc.
}
The Einstein sum convention is implied throughout this text unless otherwise specified.
The standard Euclidean scalar product on $\mathbb{R}^{3 \times 3}$ is given by $\left\langle X, Y \right\rangle= \mbox{tr} (X\cdot Y^T)$ and consequently the Frobenius tensor norm is $\lVert X \rVert^2 = \left\langle X, X\right\rangle$.
The identity tensor on $\mathbb{R}^{3\times 3}$ will be denoted by $\mathbbm{1}$; then, $\mbox{tr}(X)=\left\langle X, \mathbbm{1} \right\rangle$.
We denote by $B_L$ a bounded domain in $\mathbb{R}^3$, by $\delta B_L$ its regular boundary and by $\Sigma$ any material surface embedded in $B_L$.
The outward unit normal to $\delta  B_L$ will be denoted by $\nu$ as will the outward unit normal to a surface $\Sigma$ embedded in $B_L$.
Given a field $a$ defined on the surface $\Sigma$, we define the jump of $a$ through the surface $\Sigma$ as:
\begin{equation}
\llbracket a \rrbracket = a^+ - a^-, \qquad \text{with} \qquad a^{-} := \lim_{\substack{x \in B_L^{-}\setminus \Sigma \\ x \to \Sigma}} a ,\qquad \text{and} \qquad a^{+} := \lim_{\substack{x \in B_L^{+}\setminus \Sigma \\ x \to \Sigma}} a,
\end{equation} 
where $B_L^{-}, B_L^{+}$ are the two subdomains which result from splitting $B_L$ by the surface $\Sigma$.

Classical gradient $\nabla$ and divergence $\mbox{Div}$ operators are used throughout the paper.\!\!\!
\footnote{
	The operators $\nabla$, $\mbox{curl}$ and $\mbox{Div}$ are the classical gradient, curl and divergence operators.
	In symbols, for a field $u$ of any order, $(\nabla u)_i=u_{,i}$, for a vector field $v$, $(\mbox{curl} v)_i = \epsilon_{ijk}v_{k,j}$ and for a field $w$ of order $k>1$, $(\mbox{Div} \, w)_{i_1 i_2\ldots i_{k-1}} = w_{i_1 i_2\ldots i_k,i_k}$.
}
The subscript $,j$ indicates derivation with respect to the $j-$th component of the space variable, while the subscript $,t$ denotes derivation with respect to time.
\footnote{
	Being reserved to the time variable, the index \textit{t} is treated separately and does not comply with Einstein notation.
}

Given a time interval $[0,t_0]$, the classical macroscopic displacement field is denoted by $u(x,t)\in \mathbb{R}^3$, with $x\in B_L,\, t\in [0,t_0]$.
In the framework of enriched continuum models of the micromorphic type, extra degrees of freedom are added through the introduction of the micro-distortion tensor $P$ denoted by $P(x,t) \in \mathbb{R}^{3\times 3}$, with $x\in B_L,\, t\in [0,t_0]$.
\section{Equilibrium equations, constitutive relations, and energy flux}
\label{sec1}
\subsection{Isotropic Cauchy continuum}
\label{subsec1}
The equilibrium equations for the Cauchy continuum are
\begin{equation}
\rho \, u_{,tt} = \mbox{Div}\left[\sigma\right],
\label{eq:equiCau}
\end{equation}
where $\sigma$ is the Cauchy stress tensor.
In the isotropic case, it takes the constitutive form $\sigma = 2\mu \, \mbox{sym}\nabla u + \lambda \, \mbox{tr}\left(\mbox{sym} \nabla u\right) \mathbbm{1}$, where $\lambda$ and $\mu$ are the Lamé parameters and $\mbox{sym}\nabla u$ is the strain tensor.

When dissipative phenomena can be neglected, the following flux equation must hold:
\begin{equation}
E_{,t} + \mbox{Div} H=0 \, ,
\label{eq:EnergyConservation}
\end{equation}
where $E$ is the total energy of the system and $H$ is the energy flux vector, whose explicit expression is given by (see e.g. \cite{aivaliotis2018low} for a detailed derivation) 
\begin{equation}
H = -\sigma \cdot u_{,t} \, .
\label{eq:Cauchyflux}
\end{equation}
\subsection{Relaxed micromorphic continuum with zero static characteristic length}
\label{subsec2}
The equilibrium equations are obtained by looking for stationary points of the following action functional:
\begin{equation}
\mathcal{A} = \int_{0}^{t_{0}} \int_{B_{L}} \left(J-W\right)dXdt
\label{eq:actionFuncMic}
\end{equation}
where $J$ is the kinetic energy density and $W$ is the strain energy density of the considered micromorphic continuum.

In particular, the expression of the kinetic energy density takes the form \cite{d2019effective,romano2016micromorphic}:
\!\!\!\!\!
\footnote{
The presence of curvature terms is essential to catch size-effects in the static regime that are not the target of the present paper.
}
\begin{equation}
\begin{array}{ll}
J \left(u_{,t},\nabla u_{,t}, P_{,t}\right) =& 
\dfrac{1}{2}\rho \, \langle u_{,t},u_{,t} \rangle + 
\dfrac{1}{2} \langle \mathbb{J}_{\mbox{\tiny micro}}  \, \mbox{sym} \, P_{,t}, \mbox{sym} \, P_{,t} \rangle 
+ \dfrac{1}{2} \langle \mathbb{J}_{c} \, \mbox{skew} \, P_{,t}, \mbox{skew} \, P_{,t} \rangle \\[3mm]
& + \dfrac{1}{2} \langle \mathbb{T}_{e} \, \mbox{sym}\nabla u_{,t}, \mbox{sym}\nabla u_{,t} \rangle
  + \dfrac{1}{2} \langle \mathbb{T}_{c} \, \mbox{skew}\nabla u_{,t}, \mbox{skew}\nabla u_{,t} \rangle,
\end{array}
\label{eq:kinEneMic}
\end{equation}
where $u$ is the macroscopic displacement field, $P \in \mathbb{R}^{3\times3}$ is the non-symmetric micro-distortion tensor, $\rho$ is the macroscopic apparent density, and $\mathbb{J}_{\mbox{\tiny micro}}$, $\mathbb{J}_{c}$, $\mathbb{T}_{e}$, $\mathbb{T}_{c}$ are 4th order micro-inertia tensors whose form will be specified in the following subsection.

The relaxed micromorphic continuum contains curvature terms connected to $\mbox{Curl} P$. Here, we use the relaxed micromorphic continuum without curvature contribution.
The expression of the strain energy density without curvature contribution ($\frac{\mu \, L_c^2}{2}\lVert\mbox{Curl} P\rVert^2 = 0$, $L_c=0$) is \cite{d2019effective,romano2016micromorphic}:
\begin{equation}
\begin{array}{ll}
W \left(\nabla u, P\right) =& 
  \dfrac{1}{2} \langle \mathbb{C}_{e} \, \mbox{sym}\left(\nabla u -  \, P \right), \mbox{sym}\left(\nabla u -  \, P \right) \rangle
+ \dfrac{1}{2} \langle \mathbb{C}_{\mbox{\tiny micro}} \, \mbox{sym}  \, P,\mbox{sym}  \, P \rangle\\[3mm]
& 
+ \dfrac{1}{2} \langle \mathbb{C}_{c} \, \mbox{skew}\left(\nabla u -  \, P \right), \mbox{skew}\left(\nabla u -  \, P \right) \rangle 
\, ,
\end{array}
\label{eq:strainEneMic}
\end{equation}
where $\mathbb{C}_{e}$, $\mathbb{C}_{micro}$, and $\mathbb{C}_{c}$ are 4th order tensors whose characteristic will be given in Sec.~\ref{sec:plane_stain_symmetry}.

The minimization of the Action functional, eq.~(\ref{eq:actionFuncMic}), while using eq.~(\ref{eq:kinEneMic})-(\ref{eq:strainEneMic}) provides the following equilibrium equations
\begin{equation}
\rho \, u_{,tt} - \mbox{Div}\left(\widehat{\sigma}_{,tt}\right) = \mbox{Div}\left(\widetilde{\sigma}\right),
\qquad
\left( \mathbb{J}_{\mbox{\tiny micro}} + \mathbb{J}_{c} \right) \, P_{,tt} = \widetilde{\sigma} - s,
\label{eq:equiMic}
\end{equation}
where 
\begin{equation}
\begin{array}{cc}
\widehat{\sigma} := \mathbb{T}_{e}~\mbox{sym} \nabla u + \mathbb{T}_{c}~\mbox{skew} \nabla u ,
\qquad
s := \mathbb{C}_{\mbox{\tiny micro}}~\mbox{sym}  \, P,
\\[3mm]
\widetilde{\sigma} := \mathbb{C}_{e}~\mbox{sym}\left(\nabla u -  \, P \right) + \mathbb{C}_{c}~\mbox{skew}\left(\nabla u -  \, P \right).
\end{array}
\label{eq:sigMic}
\end{equation}
The flux equation for the relaxed micromorphic continuum is formally the same as eq.~\eqref{eq:EnergyConservation}, but $H$ has now the following expression (see \cite{aivaliotis2019relaxed} for more details):
\begin{equation}\label{eq:fluxAniso}
H = -\left(\widetilde{\sigma} + \widehat{\sigma}\right)^T\cdot u_{,t} \, .
\end{equation}
\subsection{Particularization of the relaxed micromorphic model to plane strain and tetragonal symmetry}
\label{sec:plane_stain_symmetry}
We now focus on finding solutions in a plane strain framework.
This means that we constrain the displacement field $u$ and the micro-distortion $P$ to depend only on the first two components $x_1$ and $x_2$ of the space variable $x$:

\begin{equation}
u (x_1,x_2) = 
\begin{pmatrix}
u_1(x_1,x_2)\\
u_2(x_1,x_2)\\
0\\
\end{pmatrix} \, ,
\quad
P (x_1,x_2) =  
\begin{pmatrix}
P_{11}(x_1,x_2)&P_{12}(x_1,x_2)&0\\
P_{21}(x_1,x_2)&P_{22}(x_1,x_2)&0\\
0&0&0\\
\end{pmatrix} \, .
\label{eq:plane_strain_u_u}
\end{equation}

Given the metamaterials targeted in this paper (see Section~\ref{sec:microstructure}), we particularise the equilibrium equations to the tetragonal case.
This means that the elastic and micro inertia tensors appearing in eq.~(\ref{eq:kinEneMic})-(\ref{eq:strainEneMic}) can be represented in the Voigt form as:

\begin{equation}
\begin{array}{rl}
\mathbb{C}_{e} &= 
\begin{pmatrix}
\lambda_{e} + 2\mu_{e}	& \lambda_{e}				& \dots		& \bullet\\ 
\lambda_{e}				& \lambda_{e} + 2\mu_{e}	& \dots		& \bullet\\ 
\vdots					& \vdots					& \ddots	& 		 \\ 
\bullet					& \bullet					& 			& \mu_{e}^{*}\\ 
\end{pmatrix} \, ,
\quad
\mathbb{C}_{c} = 
\begin{pmatrix}
\bullet & 			& \bullet\\ 
		& \ddots 	& \vdots\\ 
\bullet & \dots		& 4\mu_{c}
\end{pmatrix} \, ,
\\[20mm]
\mathbb{C}_{\mbox{\tiny micro}} &= 
\begin{pmatrix}
\lambda_{\tiny \mbox{micro}} + 2\mu_{\tiny \mbox{micro}}	& \lambda_{\tiny \mbox{micro}}				& \dots		& \bullet\\ 
\lambda_{\tiny \mbox{micro}}				& \lambda_{\tiny \mbox{micro}} + 2\mu_{\tiny \mbox{micro}}	& \dots		& \bullet\\ 
\vdots					& \vdots					& \ddots	&  \\ 
\bullet					& \bullet					& 			& \mu_{\tiny \mbox{micro}}^{*}\\ 
\end{pmatrix} \, ,
\end{array}
\label{eq:micro_ine_1}
\end{equation}
\vspace{0.5cm}
\begin{equation}
\begin{array}{rlrl}
	\mathbb{J}_{\mbox{\tiny micro}} &=\rho 
	\begin{pmatrix}
	L^2_{3} + 2L^{2}_{1} & L^{2}_{3}              & \dots 			& \bullet \\ 
	L^{2}_{3}            & L^{2}_{3} + 2L^{2}_{1} & \dots 			& \bullet \\ 
	\vdots               & \vdots                 & \ddots 			& \bullet \\ 
	\bullet              & \bullet                & \bullet 		& L^{*^2}_{1} \\ 
	\end{pmatrix} \, ,
	\quad
	&\mathbb{J}_{c} &= \rho
	\begin{pmatrix}
	\bullet & 			& \bullet\\ 
	& \ddots 	& \vdots\\ 
	\bullet & \dots & 4L^{2}_{2}
	\end{pmatrix} \, ,
	\\[20mm]
	\mathbb{T}_{e} &= \rho
	\begin{pmatrix}
	\overline{L}^2_{3} + 2\overline{L}^2_{1}	& \overline{L}^2_{3}        			   	& \dots		& \bullet\\ 
	\overline{L}^2_{3}         				    & \overline{L}^2_{3} + 2\overline{L}^2_{1} 	& \dots		& \bullet\\ 
	\vdots                    					& \vdots                 				   	& \ddots 	&		 \\ 
	\bullet                   					& \bullet									& 	 		& \overline{L}^{*^2}_{1}
	\end{pmatrix} \, ,
	\quad
	&\mathbb{T}_{c} &= \rho
	\begin{pmatrix}
	\bullet & 			& \bullet\\ 
	& \ddots 	& \vdots\\ 
	\bullet & \dots & 4\overline{L}^2_{2}
	\end{pmatrix} \, ,
	\end{array}
\label{eq:micro_ine_2}
\end{equation}
where only the coefficients involved in a plane strain problem are reported (the dots represent components acting on out-of plane variables and are not specified here).

In the definition (\ref{eq:micro_ine_2}) of the micro-inertia tensors appearing in the kinetic energy (\ref{eq:kinEneMic}), it is underlined the fact that they introduce dynamic internal lengths that can be directly related to the dispersion behaviour of the metamaterial at very small (in the limit vanishing) wavenumbers ($\mathbb{J}_{\mbox{\tiny micro}}$,$\mathbb{J}_{c}$), as well as at very large (in the limit infinite) wavenumbers ($\mathbb{T}_{e}$, $\mathbb{T}_{c}$).
\section{Boundary conditions for a finite-size relaxed micromorphic slab embedded between two Cauchy half-spaces}
\label{sec:BCs}
\begin{figure}[H]
	\centering
	\includegraphics[width=0.65\textwidth]{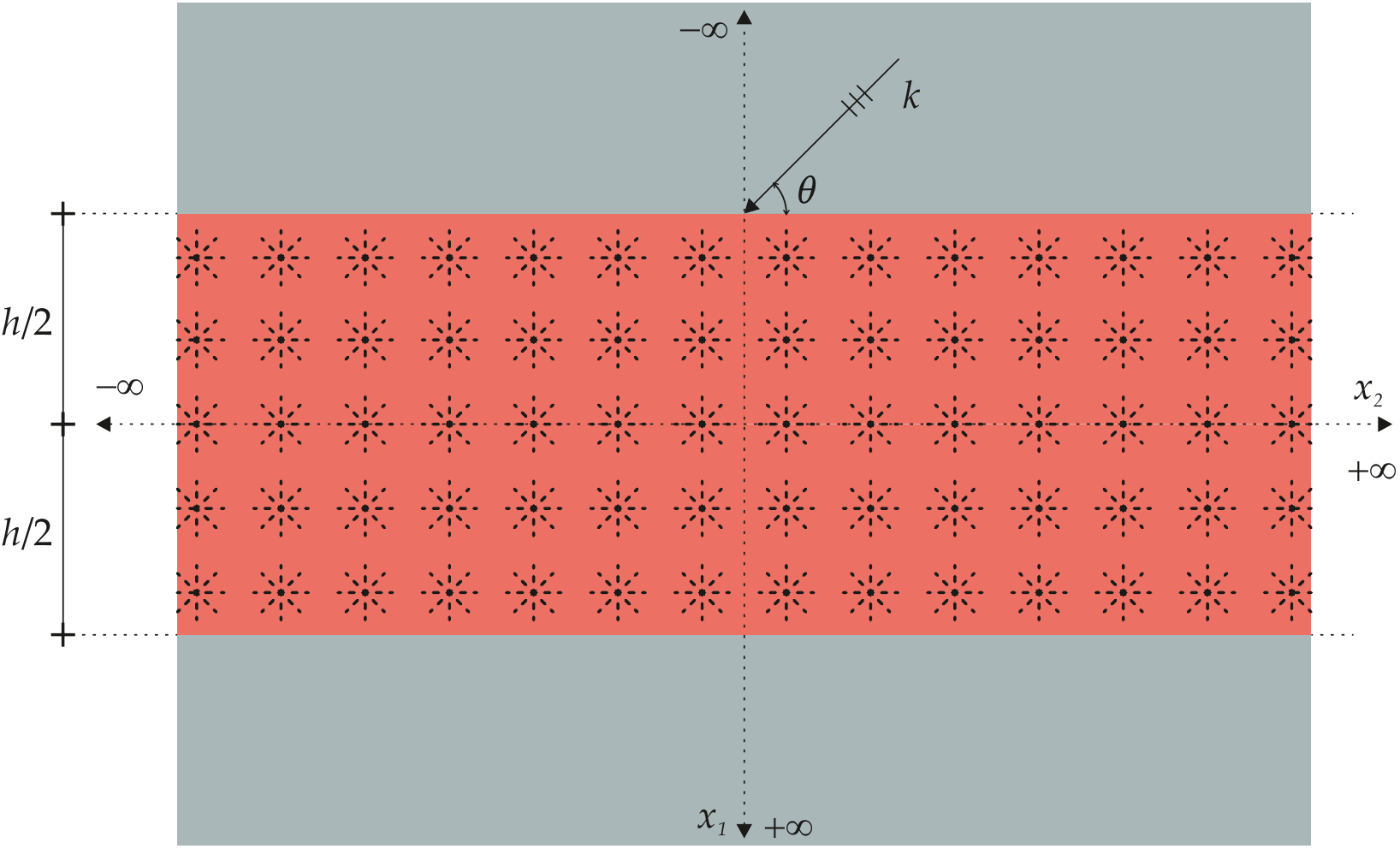}
	\caption{Schematic representation of a wave with wavenumber $k$ hitting at angle $\theta$ a relaxed micromorphic slab of thickness $h$ embedded between two isotropic Cauchy media.}
	\label{fig:slab_mic}
\end{figure}

Two half-spaces made up of a homogeneous Cauchy material are separated by a micromorphic slab of finite width $h$.
The three materials are in perfect contact with each other: the material on the top of the first interface is a classical linear elastic isotropic Cauchy medium, the material in the middle is an anisotropic relaxed micromorphic medium, while the material on the bottom of the second interface is again a classical isotropic Cauchy medium (see Fig. \ref{fig:slab_mic}).

As it can be seen in \cite{madeo2016reflection,aivaliotis2018low,aivaliotis2019microstructure} there are two boundary conditions which can be imposed at a Cauchy/relaxed-micromorphic interface if the static characteristic length ($L_c$) is zero (our case here): the continuity of displacement and continuity of generalized traction.

In the considered 2D case, there are then eight sets of scalar conditions, four on each interface. The finite slab has width $h$ and we assume that the two interfaces are positioned at $x_1=-h/2$ and $x_1=h/2$, respectively (see Fig.~\ref{fig:slab_mic}). The continuity of displacement conditions to be satisfied at the two interfaces of the slab are:
\begin{equation}\label{eq:jumpdisplslab}
u^{-}_c = u_s, \text{ on } x_1=-\frac{h}{2},
\qquad\qquad
u_s = u^{+}_c,  \text{ on } x_1=\frac{h}{2},
\end{equation}
where $u^{-}_c$ and $u^{+}_c$ are the displacement of the ``minus'' ($x_1<0$) and ``plus''($x_1>0$) Cauchy half-space, respectively.
As for the continuity of generalized traction, we have:
\begin{equation}\label{eq:jumptractionslab}
t^{-}_c = t_s,  \text{ on } x_1=-\frac{h}{2},
\qquad\qquad
t_s = t^{+}_c,   \text{ on } x_1=\frac{h}{2},
\end{equation}
where $t_{c}^{\pm} = \sigma^{\pm} \cdot \nu^{\pm}$ are classical Cauchy tractions, $t_{s} = \left(\widetilde{\sigma} + \widehat{\sigma}\right) \cdot \nu$ is the generalized traction in the relaxed micromorphic medium, with $\nu$ being the outward unit normal to the surface considered (see \cite{d2019effective,aivaliotis2020frequency} and eq.~(\ref{eq:sigMic}) for details about the definitions of generalized tractions).
\section{2D tetragonal microstructures for acoustic control}
\label{sec:microstructure}
\begin{figure}[H]
	\centering
	\includegraphics[width=0.9\textwidth]{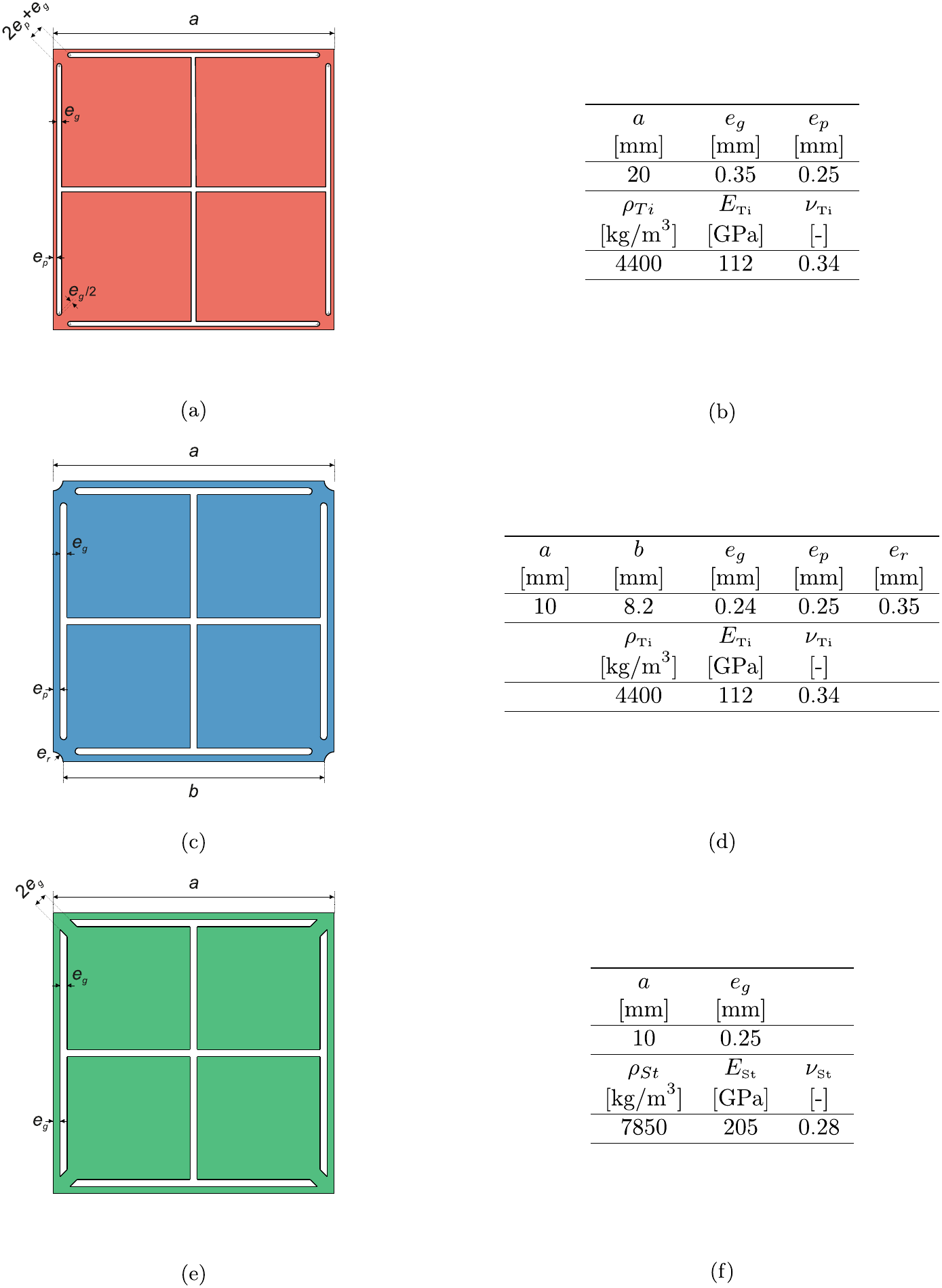}
	\caption{
		(a) unit cell giving rise to the metamaterial 1, or in short \textit{MM1}.
		(b) Table geometry and material properties of the unit cell: $\rho_{\tiny \mbox{Ti}}$, $E_{\tiny \mbox{Ti}}$, and $\nu_{\tiny \mbox{Ti}}$ stand for the density, Young modulus and Poisson's ratio of titanium, respectively.
		(c) unit cell giving rise to the metamaterial 2, or in short \textit{MM2}.
		(d) Table geometry and material properties of the unit cell: $\rho_{\tiny \mbox{Ti}}$, $E_{\tiny \mbox{Ti}}$, and $\nu_{\tiny \mbox{Ti}}$ stand for the density, Young modulus and Poisson's ratio of titanium, respectively.
		(e) unit cell giving rise to the metamaterial 3, or in short \textit{MM3}.
		(f) Table geometry and material properties of the unit cell: $\rho_{\tiny \mbox{St}}$, $E_{\tiny \mbox{St}}$, and $\nu_{\tiny \mbox{St}}$ stand for the density, Young modulus and Poisson's ratio of steel, respectively.}
	\label{fig:fig_tab_unit_cell}
\end{figure}

In view of the conception of meta-structures for applications in acoustic control, we consider here three tetragonal unit cells which give rise to three different metamaterials at the macroscopic scale.
These metamaterials will then be characterised through the relaxed micromorphic model, thus widening the set of tetragonal microstructures that have been characterised so far with this new model (see \cite{d2019effective,aivaliotis2020frequency} for the characterization of an ultrasound microstructure).
The selected microstrutures are shown in Fig.~\ref{fig:fig_tab_unit_cell}.
They all show band-gaps for relatively low frequencies, especially the one presented in Fig.~\ref{fig:fig_tab_unit_cell}(a) which completely falls in the acoustic frequencies range (see dispersion curves Fig.~\ref{fig:disp_0_tot}--\ref{fig:disp_1_2_tot}).
Given the particular distribution of voids in the unit cells, these microstructures result to be by far stiffer in compression than in shear.
This can be observed in Fig.~\ref{fig:disp_0_tot}--\ref{fig:disp_1_2_tot}, by remarking that for a wave propagating along the horizontal direction ($\theta = 0$), the acoustic ``pressure'' wave is by far steeper than the ``shear'' wave (see Fig.~\ref{fig:disp_0_tot}(a)).
This difference is lost for other directions of propagation (see, e.g., Fig.~\ref{fig:disp_0_tot}(b)--\ref{fig:disp_0_tot}(c)).

All the simulations that are reported in this section and in the next one have been carried on with the software Mathematica for what concerns the relaxed micromorphic semi-analytical solutions, and with the software Comsol for the detailed discrete numerical solutions(see \cite{aivaliotis2020frequency} for more details).
\subsection{Dispersion curves and calibration of the relaxed micromorphic elastic parameters}
In this section we provide the relaxed-micromorphic characterization of the three metamaterials previously introduced (Fig.\ref{fig:fig_tab_unit_cell}) by means of the fitting procedure developed in \cite{d2019effective,neff2019identification}.
This fitting procedure is based on two different steps aimed at separately characterizing the metamaterial on the static and dynamic regime \cite{d2019effective,neff2019identification}.
In particular, the static parameters are identified by remarking that the relaxed micromorphic model tends to a macroscopic equivalent Cauchy medium of stiffness $\lambda_{\tiny \mbox{macro}}$, $\mu_{\tiny \mbox{macro}}$, and $\mu_{\tiny \mbox{macro}}^{*}$, when considering the long-wave limit (small frequencies and small wave numbers).
These macroscopic parameters can be identified by classical numerical homogenization.
They are obtained by imposing periodic boundary conditions that mimic an infinitely extended structure \cite{d2019effective,neff2019identification}.
On the other hand, the micro-parameters $\lambda_{\tiny \mbox{micro}}$, $\mu_{\tiny \mbox{micro}}$, and $\mu_{\tiny \mbox{micro}}^{*}$ are identified by imposing kinematic uniform boundary conditions on different but equivalent and symmetry-preserving unit cells \cite{neff2019identification}.
It is important to underline that the micro-parameters so identified estimate is just a lower bound.
The parameters $\lambda_{e}$, $\mu_{e}$, $\mu^{*}_e$ are uniquely identified as a combination of the macro- and micro- elastic parameters thanks to the homogenization formulas derived in \cite{d2019effective,neff2019identification}:
\begin{equation}
\begin{array}{c}
\mu _e = \dfrac{\mu_{\tiny \mbox{macro}} \, \mu_{\tiny \mbox{micro}}}{\mu_{\tiny \mbox{micro}} - \mu_{\tiny \mbox{macro}}} \, ,
\;
\;
\mu _e^{*} = \dfrac{\mu_{\tiny \mbox{macro}}^{*} \, \mu_{\tiny \mbox{micro}}^{*}}{\mu_{\tiny \mbox{micro}}^{*} - \mu_{\tiny \mbox{macro}}^{*}} \, ,
\;
\;
\kappa _e = \dfrac{\kappa_{\tiny \mbox{macro}} \,\ \kappa_{\tiny \mbox{micro}}}{\kappa_{\tiny \mbox{micro}} - \kappa_{\tiny \mbox{macro}}} \, ,
\;
\;
\text{with}
\;
\:
\!
\left\{
\begin{array}{l}
\kappa_{\tiny \mbox{i}} = \dfrac{2\mu_{i} + 3\lambda_{i}}{3}
\\[3mm]
i = \left\{e, {\tiny \mbox{micro}}, {\tiny \mbox{macro}}\right\}
\end{array} .
\right.
\end{array}
\label{eq:static_homo_relation}
\end{equation}

As for the dynamic parameters, four of them can be computed by considering the limiting case $k \to 0$ (vanishing wavenumbers).
In particular, the parameters $L_1$, $L_2$, $L_3$, and $L_1^*$ can be determined by imposing the cut-off of the relaxed micromorphic model to be equal to the corresponding numerical values obtained, e.g., by Bloch-Floquet analysis. 
Indeed, the cut-offs frequencies of the relaxed micromorphic model are found to be given by \cite{d2019effective}:
\begin{equation}
\omega_r = \sqrt{\dfrac{\mu_c}{\rho \, L_2^2}}\, ,
\quad
\omega_s = \sqrt{\dfrac{\mu_e + \mu_{\mbox{\tiny micro}}}{\rho \, L_1^2}}\, ,
\quad
\omega_s^* = \sqrt{\dfrac{\mu_e^* + \mu_{\mbox{\tiny micro}}^*}{\rho \, L_1^{*^2}}}\, ,
\quad
\omega_p = \sqrt{\dfrac{\mu_e + \lambda_e + \mu_{\mbox{\tiny micro}} + \lambda_{\mbox{\tiny micro}}}{\rho \, \left(L_1^2 + L_3^2\right)}}\, .
\label{eq:cut-offs}
\end{equation}
The remaining dynamic parameters $\overline{L}_1$, $\overline{L}_2$,$\overline{L}_3$, and $\overline{L}_1^*$ are found to have a strong effect on the dispersion curves when $k \to \infty$ and they are determined by inverse approach to reach the best possible fitting of the Bloch-Floquet dispersion curves \cite{d2019effective,aivaliotis2020frequency}.
\subsubsection{Relaxed micromorphic characterization of the metamaterial \textit{MM1}}
As a result of the fitting procedure briefly summarized before, the metamaterial \textit{MM1} (see Fig.~\ref{fig:fig_tab_unit_cell}(a)) results to be characterized via the relaxed micromorphic parameters give in Table~\ref{tab:parameters_RM_0} (a). Table~\ref{tab:parameters_RM_0} (b) provides the corresponding values of the Cauchy medium obtained as the long wave limit of the relaxed micromorphic medium of Table~\ref{tab:parameters_RM_0} (a).
\begin{table}[H]
	\renewcommand{\arraystretch}{1.5}
	\centering
	\begin{subtable}[t]{.70\textwidth}
	\centering
		\begin{tabular}{|c|c|c|c|c|c|c|} 
			\cline{1-3}
			$\lambda_{e}$ [Pa] & $\mu_{e}$ [Pa] & $\mu^{*}_{e}$ [Pa]   \\
			\cline{1-3}
			$1.00884\times10^8$ & $2.52771\times10^9$  & $1.25592\times10^6$  \\
			\cline{1-3}
			$\lambda_{\mbox{\tiny micro}}$ [Pa] & $\mu_{\mbox{\tiny micro}}$ [Pa] & $\mu^{*}_{\mbox{\tiny micro}}$ [Pa] \\
			\cline{1-3}
			$1.832\times10^8$  & $4.50125\times10^9$  & $2.698\times10^8$  \\ 
			\hline
			$L_{1}$ [m] & $L_{2}$ [m] & $L_{3}$ [m] & $L^{*}_{1}$ [m]\\
			\hline
			$0.100$ & $1.24908\times10^{-3}$ & $2.02572\times10^{-2}$ & $2.44985\times10^{-2}$\\ 
			\hline	
			$\overline{L}_{1}$ [m] & $\overline{L}_{2}$ [m] & $\overline{L}_{3}$ [m] & $\overline{L}^{*}_{1}$ [m]\\
			\hline
			$4.5639\times10^{-4}$ & $2.28195\times10^{-3}$ & $1.44323\times10^{-3}$ & $4.84074\times10^{-3}$\\ 
			\hline	
			$\rho$ [kg/m$^3$] & $\mu_{c}$ [Pa] \\
			\cline{1-2}
			$3841$ & $10^5$  \\ 
			\cline{1-2}
		\end{tabular}
	\subcaption{}
	\end{subtable}
	\hfill
	\centering
	\begin{subtable}[t]{.20\textwidth}
	\centering
		\begin{tabular}{|c|c|c|} 
			\hline
			$\lambda_{\tiny \mbox{macro}}$ [Pa]\\
			\hline
			$6.507\times10^7$\\
			\hline
			$\mu_{\tiny \mbox{macro}}$ [Pa]\\
			\hline
			$1.619\times10^9$\\
			\hline
			$\mu^{*}_{\tiny \mbox{macro}}$ [Pa]\\
			\hline
			$1.250\times10^6$\\
			\hline
		\end{tabular}
	\vspace{1.1cm}
	\subcaption{}
	\end{subtable}
\caption{
	Panel (a) shows the values of the relaxed micromorphic static and dynamic parameters for the metamaterial \textit{MM1} determined via the fitting procedure given in \protect\cite{d2019effective,aivaliotis2020frequency}. The apparent density $\rho$ is computed based on the titanium microstructure of Fig.~\ref{fig:fig_tab_unit_cell}(a).
	Panel (b) shows the values of the equivalent Cauchy continuum elastic coefficients corresponding to the long-wave limit of \textit{MM1} computed with the procedure explained in \protect\cite{neff2019identification}.
}
\label{tab:parameters_RM_0}
\end{table}
\begin{figure}[H]
	\centering
	\includegraphics[width=0.8\textwidth]{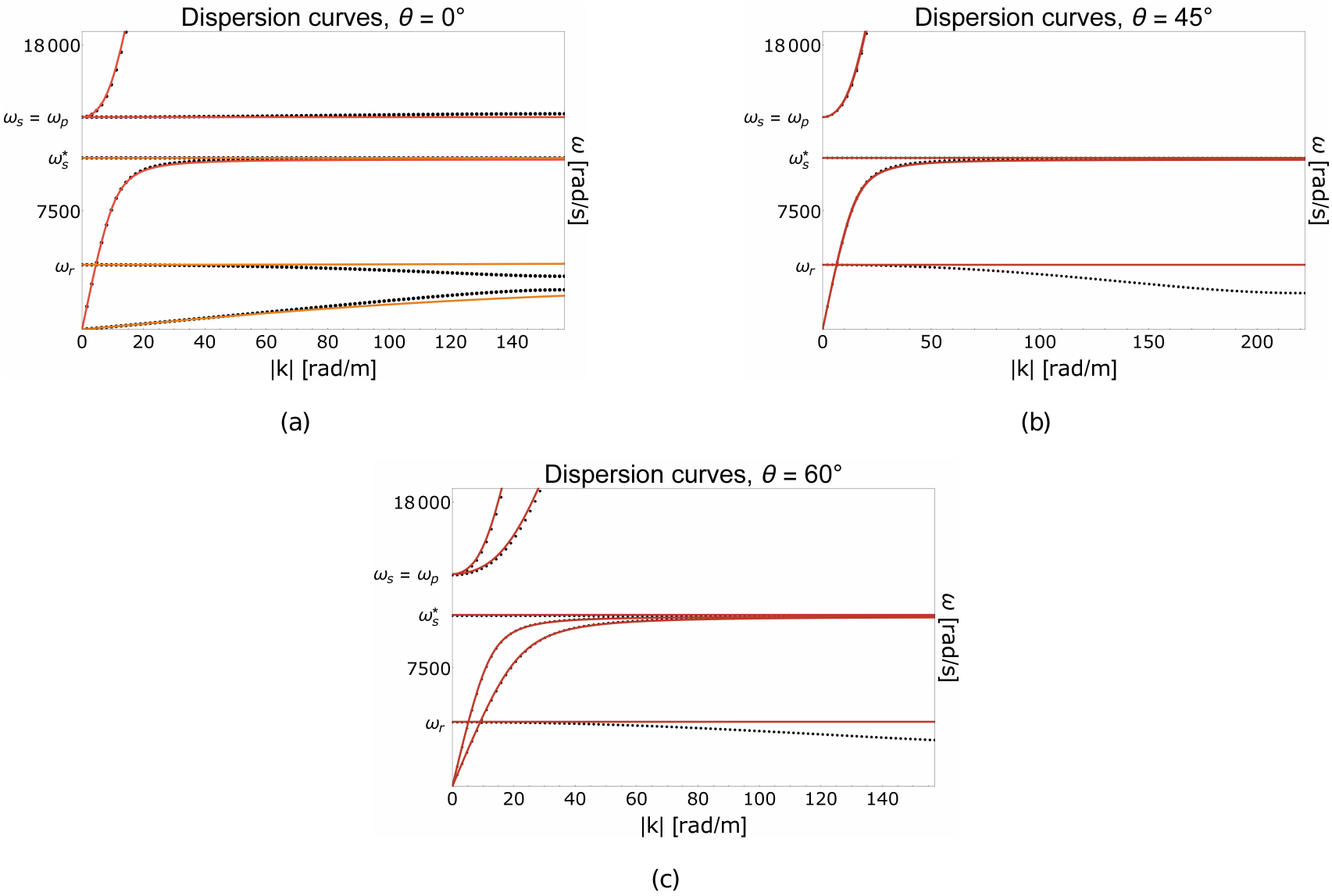}
	\caption{Dispersion curves for the metamaterial \textit{MM1} corresponding to three different direction of propagation (a) $\theta = 0^{\circ}$, (b) $\theta = 45^{\circ}$, and (c) $\theta = 60^{\circ}$.}
	\label{fig:disp_0_tot}
\end{figure}
Figure~\ref{fig:disp_0_tot} shows the comparison of the dispersion curves obtained via the relaxed micromorphic model with those issued via Bloch-Floquet analysis, for three different directions of propagation.
\subsubsection{Relaxed micromorphic characterization of the metamaterials \textit{MM2} and \textit{MM3}}
\begin{figure}[H]
	\centering
	\includegraphics[width=0.8\textwidth]{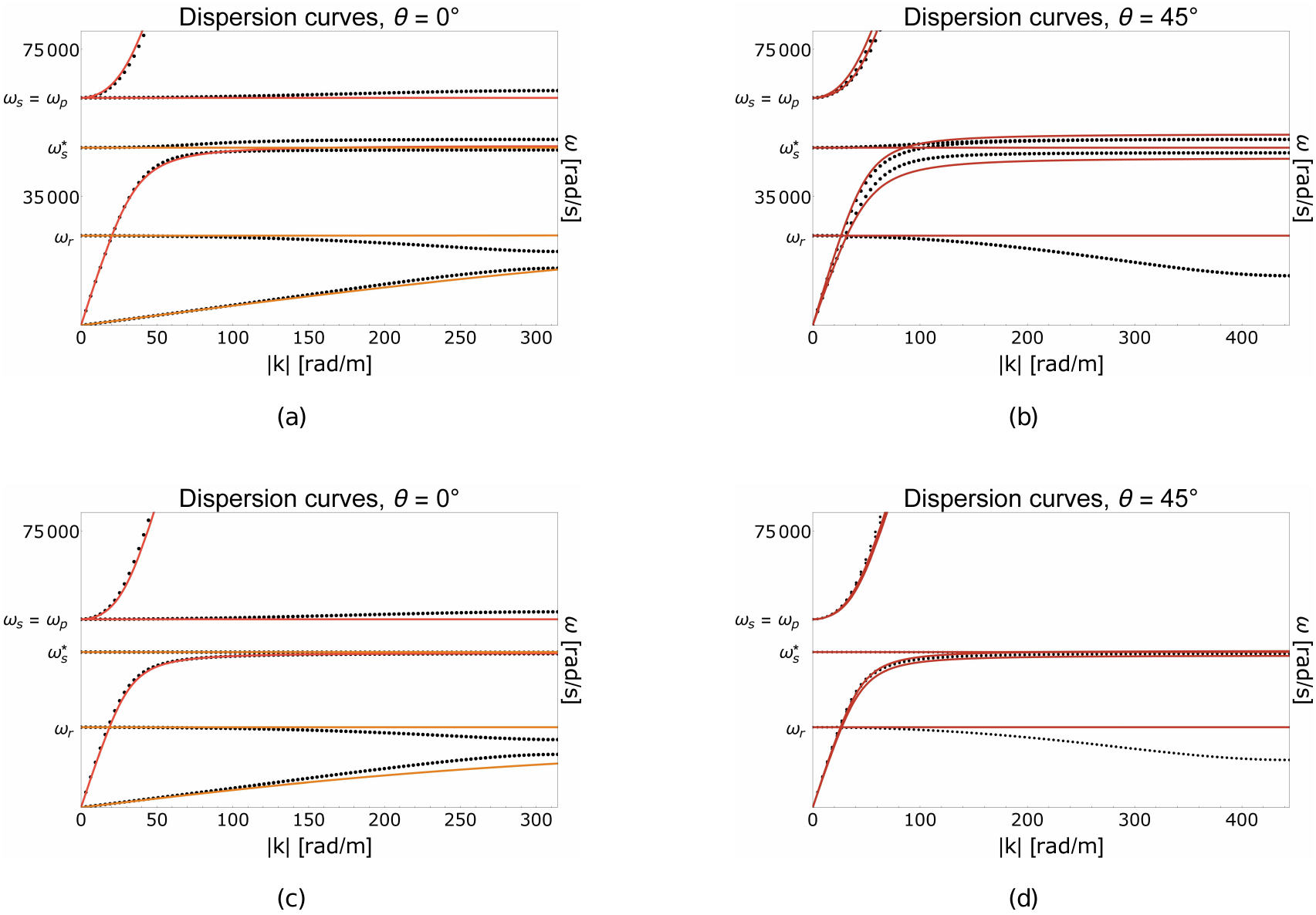}
	\caption{Dispersion curves for the metamaterial \textit{MM2} (a)-(b) and \textit{MM3}  (c)-(d) corresponding to two direction of propagation (a)-(c) $\theta = 0^{\circ}$ and (b)-(d) $\theta = 45^{\circ}$.}
	\label{fig:disp_1_2_tot}
\end{figure}

Following the aforementioned procedure, the metamaterials \textit{MM2} and \textit{MM3} are characterized through the relaxed micromorphic parameters given in Table~\ref{tab:parameters_RM_1} and Table~\ref{tab:parameters_RM_2}, respectively.
The resulting fitting of the dispersion curves is shown in Fig.~\ref{fig:disp_1_2_tot}.
\begin{table}[H]
	\renewcommand{\arraystretch}{1.5}
	\centering
	\begin{subtable}[t]{.70\textwidth}
		\centering
		\begin{tabular}{|c|c|c|c|c|c|c|} 
			\cline{1-3}
			$\lambda_{e}$ [Pa] & $\mu_{e}$ [Pa] & $\mu^{*}_{e}$ [Pa] \\
			\cline{1-3}
			$6.35565\times10^7$ & $4.96092\times10^9$  & $1.13113\times10^7$  \\
			\cline{1-3}
			$\lambda_{\mbox{\tiny micro}}$ [Pa] & $\mu_{\mbox{\tiny micro}}$ [Pa] & $\mu^{*}_{\mbox{\tiny micro}}$ [Pa]\\
			\cline{1-3}
			$6.553\times10^9$  & $5.9\times10^9$  & $5.984\times10^9$  \\ 
			\hline
			$L_{1}$ [m] & $L_{2}$ [m] & $L_{3}$ [m] & $L^{*}_{1}$ [m]\\
			\hline
			$0.0272027$ & $0.000209403$ & $0.021232$ & $0.0258829$\\ 
			\hline	
			$\overline{L}_{1}$ [m] & $\overline{L}_{2}$ [m] & $\overline{L}_{3}$ [m] & $\overline{L}^{*}_{1}$ [m]\\
			\hline
			$0.0000368211$ & $0$ & $0.0000368211$ & $0.0000260365$\\ 
			\hline	
			$\rho$ [kg/m$^3$] & $\mu_{c}$ [Pa]\\
			\cline{1-2}
			$3840.77$ & $100000$ \\ 
			\cline{1-2}
		\end{tabular}
		\subcaption{}
	\end{subtable}
	\hfill
	\centering
	\begin{subtable}[t]{.20\textwidth}
		\centering
		\begin{tabular}{|c|c|c|} 
			\hline
			$\lambda_{\tiny \mbox{macro}}$ [Pa]\\
			\hline
			$7.5424\times10^8$\\
			\hline
			$\mu_{\tiny \mbox{macro}}$ [Pa]\\
			\hline
			$2.69493\times10^9$\\
			\hline
			$\mu^{*}_{\tiny \mbox{macro}}$ [Pa]\\
			\hline
			$1.129\times10^7$\\
			\hline
		\end{tabular}
		\vspace{1.1cm}
		\subcaption{}
	\end{subtable}
	\caption{
		Panel (a) shows the values of the relaxed micromorphic static and dynamic parameters for the metamaterial \textit{MM2} determined via the fitting procedures given in \protect\cite{d2019effective,aivaliotis2020frequency}. The apparent density $\rho$ is computed based on the titanium microstructure of Fig.~\ref{fig:fig_tab_unit_cell}(c).
		Panel (b) shows the values of the equivalent Cauchy continuum elastic coefficients corresponding to the long-wave limit of \textit{MM2} computed with the procedure explained in \protect\cite{neff2019identification}.
	}
	\label{tab:parameters_RM_1}
\end{table}
\begin{table}[H]
	\renewcommand{\arraystretch}{1.5}
	\centering
	\begin{subtable}[t]{.70\textwidth}
		\centering
		\begin{tabular}{|c|c|c|c|c|c|c|} 
			\cline{1-3}
			$\lambda_{e}$ [Pa] & $\mu_{e}$ [Pa] & $\mu^{*}_{e}$ [Pa] \\
			\cline{1-3}
			$1.96607 \times 10^8$ & $8.37313 \times 10^9$  & $1.80302\times10^7$  \\ 
			\cline{1-3}
			$\lambda_{\mbox{\tiny micro}}$ [Pa] & $\mu_{\mbox{\tiny micro}}$ [Pa] & $\mu^{*}_{\mbox{\tiny micro}}$ [Pa]\\
			\cline{1-3}
			$2.5\times10^9$  & $17\times10^9$  & $1.075\times10^10$  \\ 
			\hline
			$L_{1}$ [m] & $L_{2}$ [m] & $L_{3}$ [m] & $L^{*}_{1}$ [m]\\
			\hline
			$0.0353782$ & $3.62851 \times 10^{-7}$ & $0.0129766$ & $0.028261$\\ 
			\hline	
			$\overline{L}_{1}$ [m] & $\overline{L}_{2}$ [m] & $\overline{L}_{3}$ [m] & $\overline{L}^{*}_{1}$ [m]\\
			\hline
			$9.30984 \times 10^{-8}$ & $ 4.16349\times 10^{-8}$ & $ 9.30984\times 10^{-8}$ & $0.0000294403$\\ 
			\hline	
			$\rho$ [kg/m$^3$] & $\mu_{c}$ [Pa]\\
			\cline{1-2}
			$7595.26$ & $0.473178$ \\ 
			\cline{1-2}
		\end{tabular}
		\subcaption{}
	\end{subtable}
	\hfill
	\centering
	\begin{subtable}[t]{.20\textwidth}
		\centering
		\begin{tabular}{|c|c|c|} 
			\hline
			$\lambda_{\tiny \mbox{macro}}$ [Pa]\\
			\hline
			$3.36\times10^8$\\
			\hline
			$\mu_{\tiny \mbox{macro}}$ [Pa]\\
			\hline
			$5.61\times10^9$\\
			\hline
			$\mu^{*}_{\tiny \mbox{macro}}$ [Pa]\\
			\hline
			$1.8\times10^7$\\
			\hline
		\end{tabular}
		\vspace{1.1cm}
		\subcaption{}
	\end{subtable}
	\caption{
		Panel (a) shows the values of the relaxed micromorphic static and dynamic parameters for the metamaterial \textit{MM3} determined via the fitting procedures given in \protect\cite{d2019effective,aivaliotis2020frequency}. The apparent density $\rho$ is computed based on the steel microstructure of Fig.~\ref{fig:fig_tab_unit_cell}(e).
		Panel (b) shows the values of the equivalent Cauchy continuum elastic coefficients corresponding to the long-wave limit of \textit{MM3} computed with the procedure explained in \protect\cite{neff2019identification}.
	}
	\label{tab:parameters_RM_2}
\end{table}
\section{Metastructure's refractive behaviour}
\begin{figure}[H]
	\centering
	\includegraphics[width=0.65\textwidth]{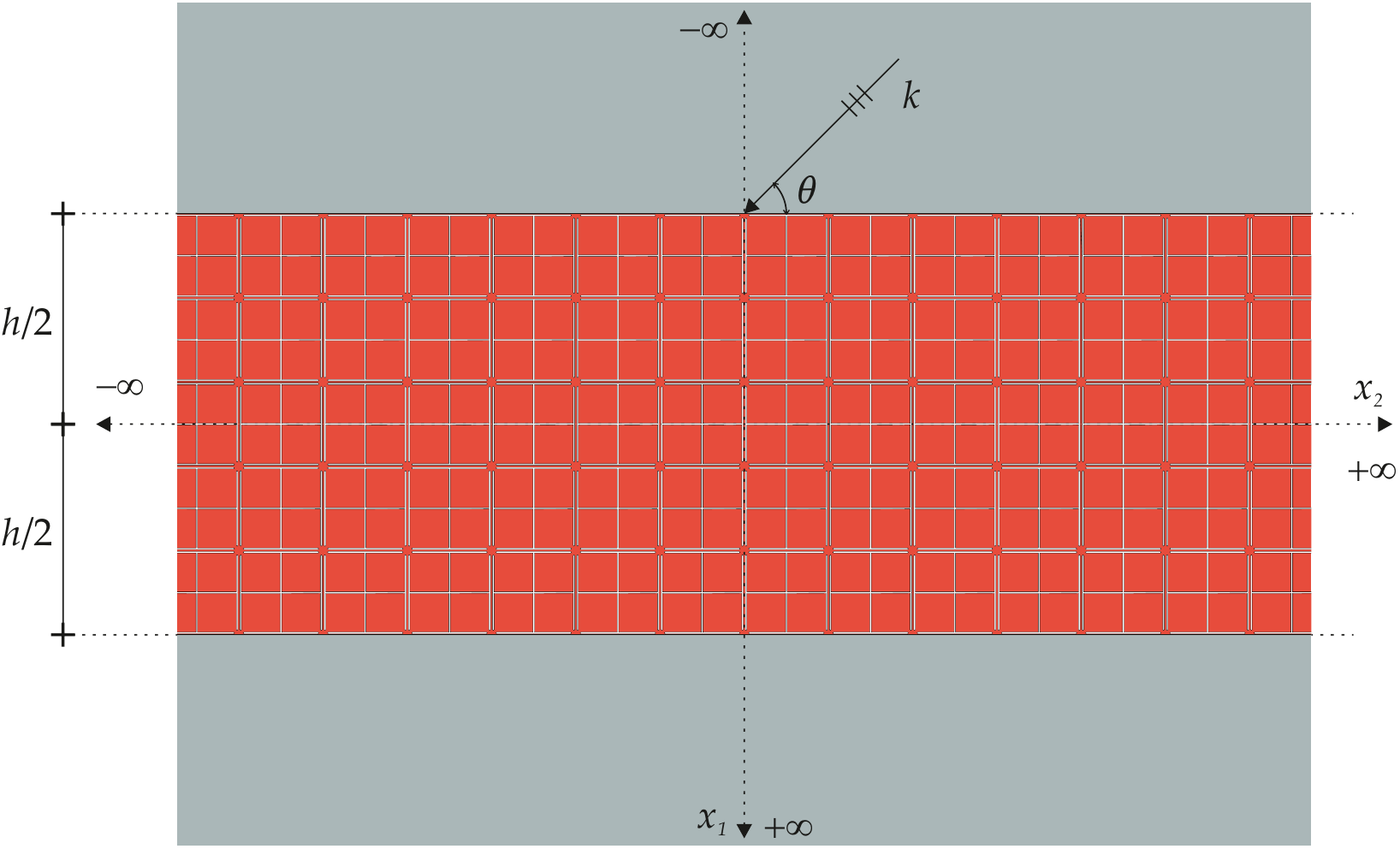}
	\caption{Schematic representation of a wave with wavenumber $k$ hitting at angle $\theta$ a microstructured material slab of thickness $h$ embedded between two isotropic Cauchy media.}
	\label{fig:slab_stru}
\end{figure}

In this section we will show how the relaxed micromorphic model can be suitably used to describe the reflective properties of a metamaterial's slab embedded in a homogeneous material (see Fig.~\ref{fig:slab_stru}).
In what follows, we will restrict ourselves to the microstructure defined in Fig.~\ref{fig:fig_tab_unit_cell}(a), since the results for the other microstructures are analogous.

We will start by considering the simpler case in which the external homogeneous material is the same as the one used for the metamaterial \textit{MM1} (see Fig.~\ref{fig:fig_tab_unit_cell}(a) for its elastic characteristics) and we will then explore how the reflective metamaterial's behaviour changes when changing the outer Cauchy material properties.
As we will see, this will allow us to explore the effect of the wavelength of the incident wave on the performances of the relaxed micromorphic model.
We will then study how the meta-structure's behaviour as well as the performances of the relaxed micromorphic model vary when increasing the number of unit cells in the metamaterial's slab.
\subsection{Dependence of the metastructure's reflective behaviour on the elastic properties of the outer Cauchy materials}
\label{subsec:reflect_elastc_properties_outer}
\begin{figure}[H]
	\centering
	\includegraphics[width=0.8\textwidth]{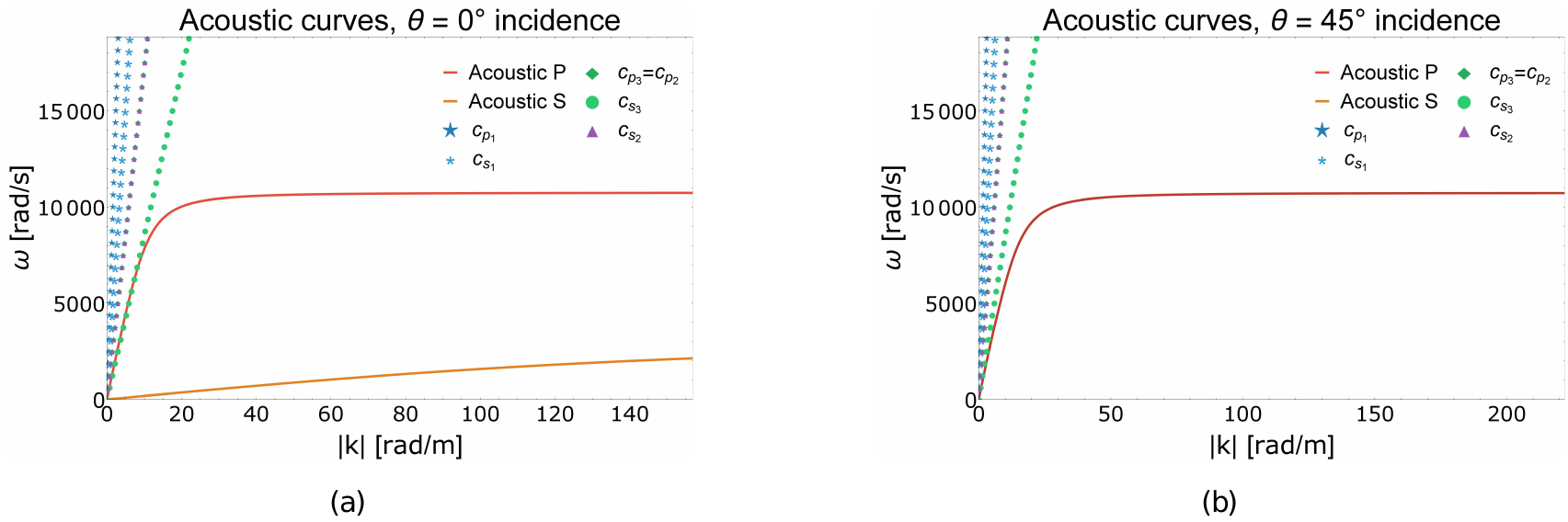}
	\caption{Acoustic curves of the metamaterial \textit{MM1} (continuous lines) as compared to the dispersion curves of three different Cauchy materials (dotted lines).
	It is highlighted that two dispersion curves coincide since $c_{p_2} = c_{p_3}$, while a third one is almost superimposed since its value $c_{s_2}$ is close to $c_{p_2}$ and $c_{p_3}$ (see Tables~\ref{tab:wave_speed_ti}-\ref{tab:wave_speed_soft_ti}-\ref{tab:wave_speed_negative_poisson} for numerical values of these speeds).}
	\label{fig:acoust_curves_0}
\end{figure}

We will show in this subsection that the refractive behaviour of the used metamaterial can be drastically modified by simply changing the elastic properties of the outer Cauchy materials.
More precisely, the simple fact of changing the properties of the external material can actually ``reverse" the metamaterial slab's refractive behaviour (from total reflection to total transmission and vice-versa).
This drastic change can be engineered for an extended range of frequencies and angles of incidence.
In order to drive the exploration of the more performing structures, we took advantage of the computational performance of the relaxed micromorphic model that allowed us to test different structures in an otherwise unreachable limited time.
We show in Fig.~\ref{fig:acoust_curves_0} the acoustic dispersion curves of the internal metamaterial, as compared to those of three different ``outer'' Cauchy materials.

\begin{figure}[H]
	\centering
	\includegraphics[width=0.8\textwidth]{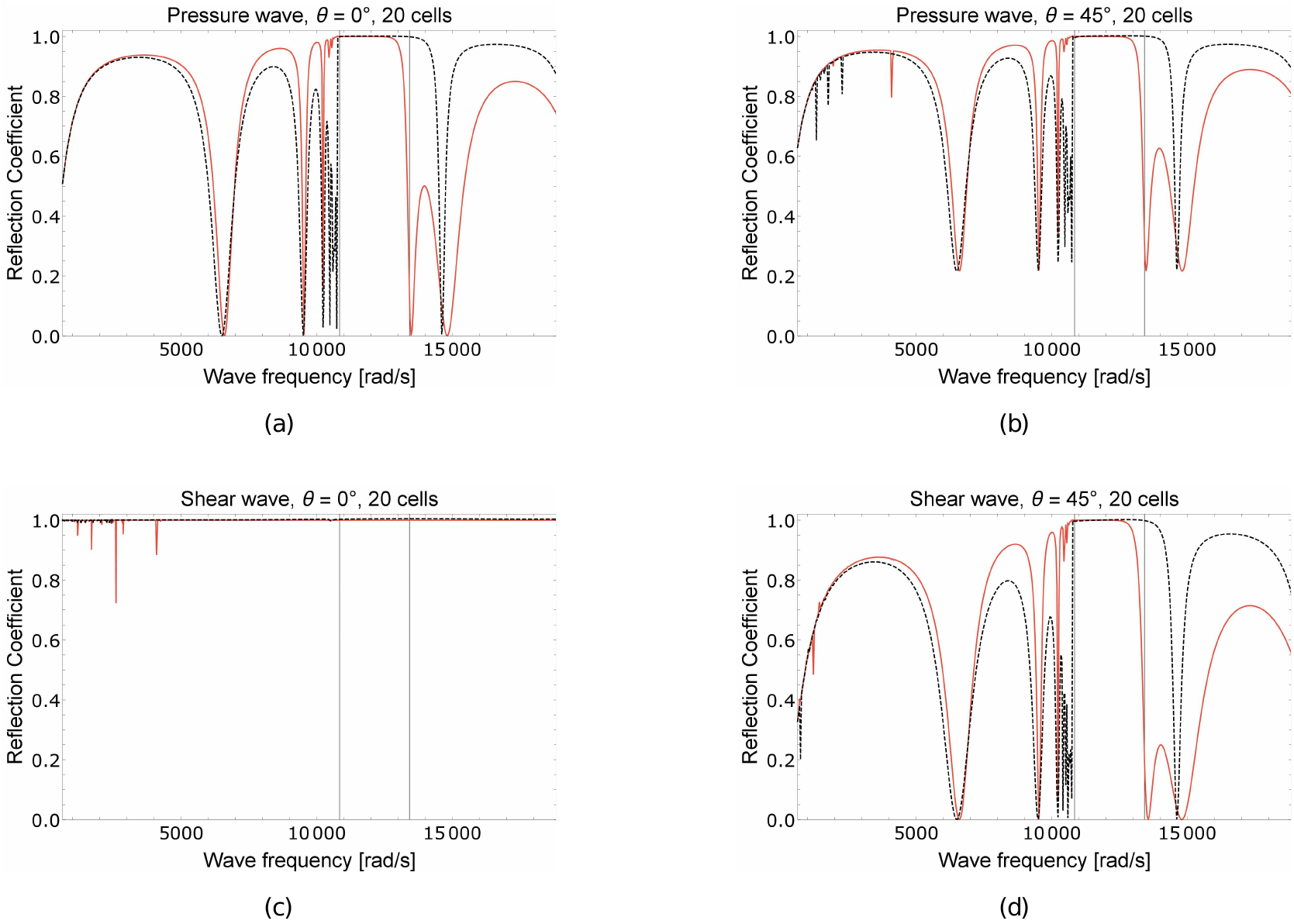}
	\caption{Comparison of the microstructure's (black line) and micromorphic (red line) reflection coefficient as a function of frequency for a 20 unit cells slab of \textit{MM1} and embedded in \textit{CM1}.
	(a) ``pressure'' normal incident wave with respect to the slab's interface.
	(b) ``pressure'' 45$^{\circ}$ incident wave with respect to the slab's interface.
	(c) ``shear'' incident wave normal to the slab's interface.
	(d) ``shear'' 45$^{\circ}$ incident wave with respect to the slab's interface.}
\label{fig:20_cell}
\end{figure}

\begin{figure}[H]
	\centering
	\includegraphics[width=0.8\textwidth]{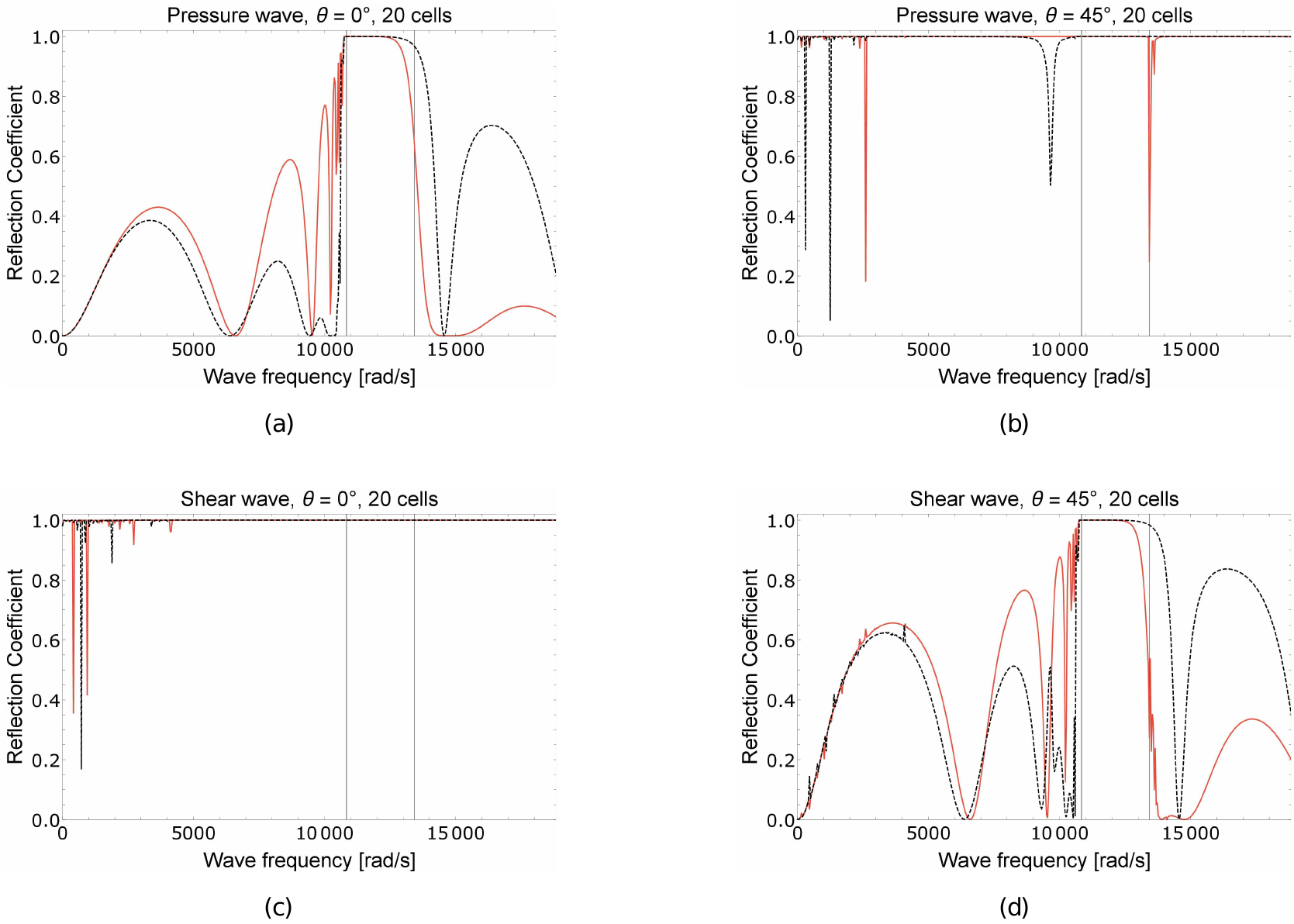}
	\caption{Comparison of the microstructure's (black line) and micromorphic (red line) reflection coefficient as a function of frequency for a 20 unit cells slab of \textit{MM1} and embedded in \textit{CM2}.
	(a) ``pressure'' normal incident wave with respect to the slab's interface.
	(b) ``pressure'' 45$^{\circ}$ incident wave with respect to the slab's interface.
	(c) ``shear'' incident wave normal to the slab's interface.
	(d) ``shear'' 45$^{\circ}$ incident wave with respect to the slab's interface.}
\label{fig:20_cell_P_-1}
\end{figure}

\begin{figure}[H]
	\centering
	\includegraphics[width=0.8\textwidth]{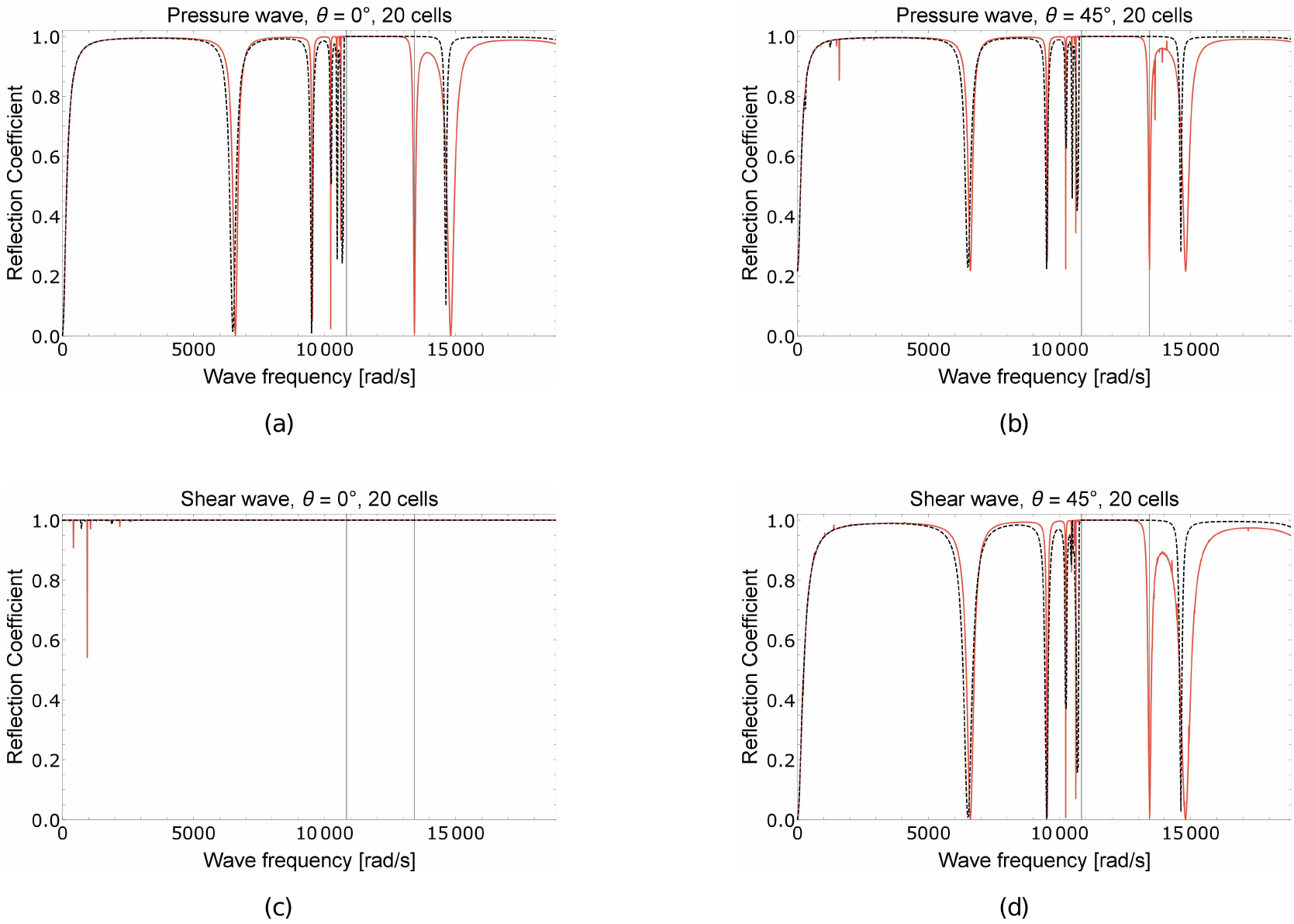}
	\caption{Comparison of the microstructure's (black line) and micromorphic (red line) reflection coefficient as a function of frequency for a 20 unit cells slab of \textit{MM1} and embedded in \textit{CM3}.
	(a) ``pressure'' normal incident wave with respect to the slab's interface.
	(b) ``pressure'' 45$^{\circ}$ incident wave with respect to the slab's interface.
	(c) ``shear'' incident wave normal to the slab's interface.
	(d) ``shear'' 45$^{\circ}$ incident wave with respect to the slab's interface.}
\label{fig:20_cell_SW}
\end{figure}

We can see from Fig.~\ref{fig:acoust_curves_0} that, in all considered cases, the external Cauchy material is relatively stiffer than the internal metamaterial in the long-wave limit, especially with the reference to the shear modulus.
This can be inferred by recalling that the slope of the acoustic curves at the origin represents the speed of propagation ($c_{\tiny \mbox{p}}$ and $c_{\tiny \mbox{s}}$) of the corresponding waves (see Table~\ref{tab:speed_macro_RM_0}) and that these speeds are directly related to the metamaterial's elastic properties in the long wave limit (see Table~\ref{tab:parameters_RM_0}).
\begin{table}[H]
	\renewcommand{\arraystretch}{2.5}
	\centering
	\begin{tabular}{|c|c|c|c|}
		\cline{1-2} \cline{3-4} 
		\multicolumn{4}{|c|}{Macro wave speed of \textit{MM1} [m/s]} \\
		\hline
		$c_p^{0} = \sqrt{\dfrac{\lambda_{\tiny \mbox{macro}} + 2\mu_{\tiny \mbox{macro}}}{\rho_{\tiny \mbox{macro}}}}$   & $c_s^{0} = \sqrt{\dfrac{\mu_{\tiny \mbox{macro}}^{*}}{\rho_{\tiny \mbox{macro}}}}$  & $c_p^{45} = \sqrt{\dfrac{\lambda_{\tiny \mbox{macro}} + \mu_{\tiny \mbox{macro}} + \mu_{\tiny \mbox{macro}}^{*}}{\rho_{\tiny \mbox{macro}}}}$  & $c_s^{45} = \sqrt{\dfrac{\mu_{\tiny \mbox{macro}}}{\rho_{\tiny \mbox{macro}}}}$    \\ \hline
		927.28 [m/s]      & 18.04 [m/s]     & 662.36 [m/s]       & 649.20 [m/s]        \\ 
		\hline
	\end{tabular}
	\caption{Wave speed expressions for the Cauchy material which is the long-wave limit of \textit{MM1}. The expressions for the pressure and shear waves are explicitly given for the two directions of propagation $\theta = 0$ and $\theta = \pi/4$. Such speeds are the slopes of the tangents at the origin of the acoustic dispersion curves in Fig.~\ref{fig:disp_0_tot}(a) and Fig.~\ref{fig:disp_0_tot}(b) and are computed based on the macro parameters given in the table in Fig.~\ref{fig:fig_tab_unit_cell}(a).}
	\label{tab:speed_macro_RM_0}
\end{table}
\begin{table}[H]
	\renewcommand{\arraystretch}{2.25}
	\begin{minipage}[t]{0.41\textwidth}
		\centering
		\begin{tabular}{|c|c|c|c|c|c|c|c|}
			\cline{1-2}
			\multicolumn{2}{|c|}{Wave speed of \textit{CM1} (Titanium) [m/s]}
			\\ 
			\cline{1-2} 
			$c_{p_{1}} = \sqrt{\dfrac{\lambda_{\tiny \mbox{Ti}} + 2\mu_{\tiny \mbox{Ti}}}{\rho_{\tiny \mbox{Ti}}}}$ & $c_{s_{1}} = \sqrt{\dfrac{\mu_{\tiny \mbox{Ti}}}{\rho_{\tiny \mbox{Ti}}}}$
			\\ 
			\cline{1-2} 
			6259.18                                     & 3081.84                          
			\\ \cline{1-2} 
		\end{tabular}
		\caption{Wave speed expressions for the Cauchy material \textit{CM1}.}
		\label{tab:wave_speed_ti}
	\end{minipage}
	\hfill
	\begin{minipage}[t]{0.41\textwidth}
		\centering
		\begin{tabular}{|c|c|c|c|c|c|c|c|}
			\cline{1-2} 
			\multicolumn{2}{|c|}{Wave speed of \textit{CM2} [m/s]}
			\\ 
			\cline{1-2}
			$c_{p_{2}} = \sqrt{\dfrac{\lambda_{\tiny \mbox{Neg}} + 2\mu_{\tiny \mbox{Neg}}}{\rho_{\tiny \mbox{Ti}}}}$ & 
			$c_{s_{2}} = \sqrt{\dfrac{\mu_{\tiny \mbox{Neg}}}{\rho_{\tiny \mbox{Ti}}}}$
			\\ 
			\cline{1-2}
			1735.98                                       & 1730                          
			\\ \cline{1-2}
		\end{tabular}
		\caption{Wave speed expressions for the Cauchy material \textit{CM2}.}
		\label{tab:wave_speed_negative_poisson}
	\end{minipage}
	\\[0.5cm]
	\centering
	\begin{minipage}[t]{0.41\textwidth}
		\centering
		\begin{tabular}{|c|c|c|c|c|c|c|c|}
			\cline{1-2} 
			\multicolumn{2}{|c|}{Wave speed of \textit{CM3} [m/s]}
			\\ 
			\cline{1-2}
			$c_{p_{2}} = \sqrt{\dfrac{\lambda_{\tiny \mbox{Ti}} + 2\mu_{\tiny \mbox{Ti}}}{13\rho_{\tiny \mbox{Ti}}}}$ & 
			$c_{p_{3}} = \sqrt{\dfrac{\mu_{\tiny \mbox{Ti}}}{13\rho_{\tiny \mbox{Ti}}}}$
			\\ 
			\cline{1-2}
			1735.98                                     & 854.75                          
			\\ \cline{1-2}
		\end{tabular}
		\caption{Wave speed expressions for the Cauchy material \textit{CM3}.}
		\label{tab:wave_speed_soft_ti}
	\end{minipage}
	\label{tab:wave_speed_cauchy}
\end{table}
The three outer Cauchy materials have been chosen starting from the same material as the one constituting the metamaterial \textit{MM1} (titanium) and then lowering the propagation speeds $c_{\tiny \mbox{p}}$ and $c_{\tiny \mbox{s}}$ so as to widen the range of frequencies for which the relaxed micromorphic model gives quantitatively good results.
\footnote{
We lowered the propagation speeds $c_{\tiny \mbox{p}}$ and $c_{\tiny \mbox{s}}$ by directly changing the values of the the stiffness of the outer metamaterial or equivalently by increasing the density as shown in Tables~\ref{tab:wave_speed_ti}-\ref{tab:wave_speed_negative_poisson}-\ref{tab:wave_speed_soft_ti} so as to meet these propagation speeds. This means that different materials \textit{CM1}, \textit{CM2}, and \textit{CM3} can be found that have the same speeds as in Tables~\ref{tab:wave_speed_ti}-\ref{tab:wave_speed_negative_poisson}-\ref{tab:wave_speed_soft_ti}.
All these materials would give rise to the meta-structure's behaviours presented in this section.
}
Indeed, with reference to Figs.~\ref{fig:20_cell}-\ref{fig:20_cell_P_-1}-\ref{fig:20_cell_SW}, we can remark that the frequency interval for which the relaxed micromorphic model is predictive of the microstructure's reflective behaviour is larger for the ``softer'' outer Cauchy material \textit{CM3} (Fig.~\ref{fig:20_cell_SW}).
Indeed, the fact of considering a ``softer'' outer material is equivalent to say that, at any given frequency, the corresponding wavelength of the travelling incident wave is smaller than the one of the wave propagating in the ``stiffer'' material.
It is known that, as far as a homogenized model is concerned, its accuracy for the study of a problem of the type presented in this paper, may depend on three different characteristic lengths, namely:
\begin{itemize}
	\item the wavelength of the travelling incident wave;
	\item the thickness of the metamaterials's slab;
	\item the characteristic size of the unit cell.
\end{itemize}
Having fixed the unit cell's dimensions for applications in acoustic control, we do not discuss here the influence of the third characteristic length.
As for the influence of the thickness of the metamaterial's slab, we refer to Section~\ref{sec:refle_curves}.

Here, we limit ourselves to remark that, given the intrinsic simplifications associated to a continuum model, a threshold value $\ell_t$ for the wavelength of the incident wave exists, below which the model starts loosing its predictive capabilities.
For the meta-structures of Fig.~\ref{fig:20_cell}-\ref{fig:20_cell_P_-1} this threshold value is reached already for frequencies slightly higher than the metamaterial's band-gap.
When considering the ``softer'' outer Cauchy material \textit{CM3}, the wavelength of the incident wave remains lower than the threshold $\ell_t$ for a larger frequency range that  exceeds the band-gap (see Fig.~\ref{fig:20_cell_SW}).
Being aware of the existence of such threshold $\ell_t$ is essential for a correct use of enriched continuum models over the appropriate frequency ranges. 
\subsection{Dependence of the meta-material's reflective behaviour on the thickness of the slab}
\label{sec:refle_curves}
In this subsection, we show to which extent the reflective behaviour of the metamaterial's slab is influenced by the thickness of the slab itself.
At the same time, we are able to show that the performances of the relaxed micromorphic model increases when increasing the thickness of the slab.

\begin{figure}[H]
	\centering
	\includegraphics[width=0.8\textwidth]{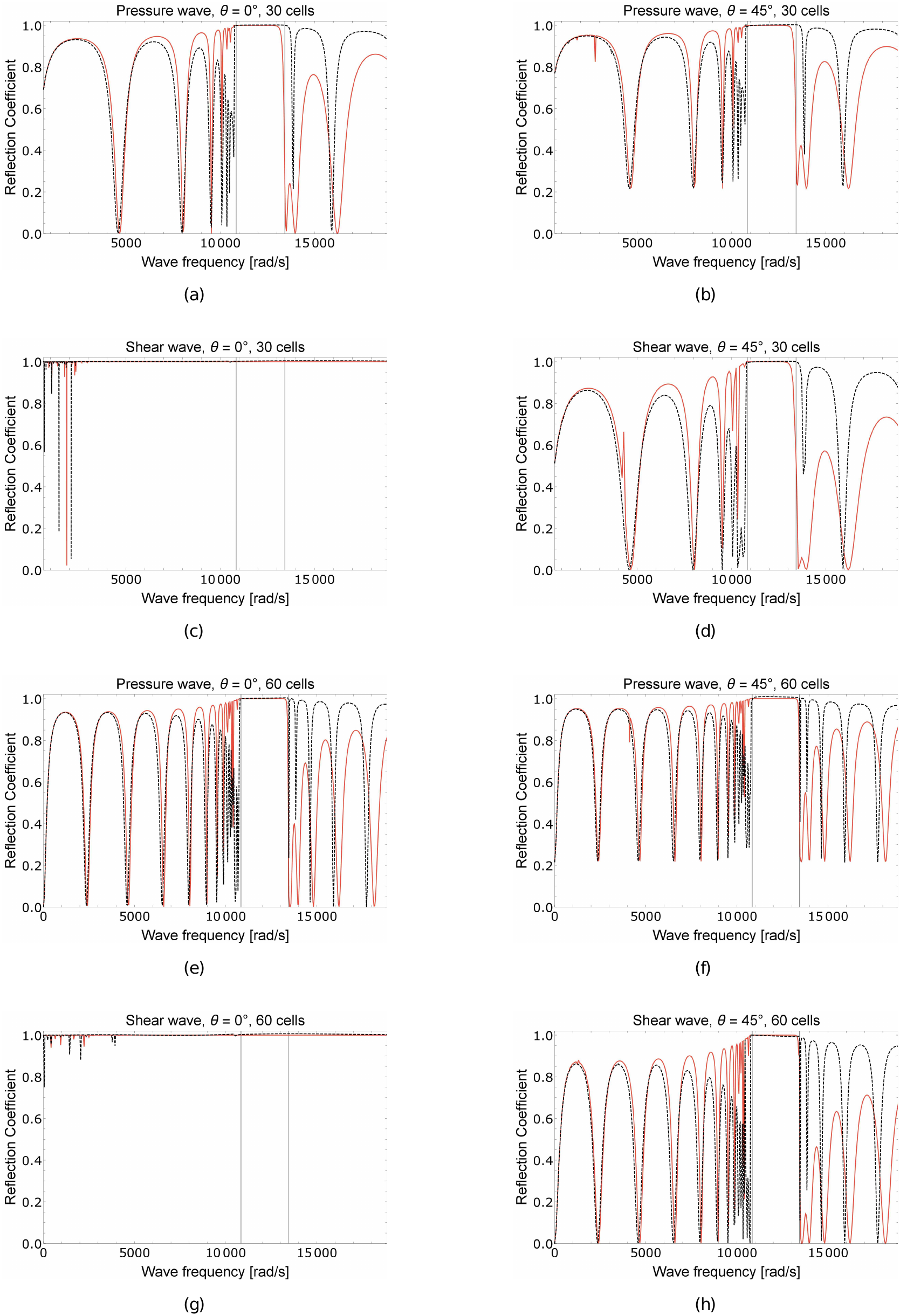}
	\caption{
	Comparison of the microstructure's (dashed black line) and micromorphic (red line) reflection coefficient as a function of frequency for a 30 (a)-(b)-(c)-(d) and a 60 (e)-(f)-(g)-(h) unit cells slab of \textit{MM1} and embedded in \textit{CM1}.
	(a) and (e) ``pressure'' normal incident wave with respect to the slab's interface for a 25 and 30 cells respectively.
	(b) and (f) ``pressure'' 45$^{\circ}$ incident wave with respect to the slab's interface for a 25 and 30 cells respectively.
	(c) and (g) ``shear'' incident wave normal to the slab's interface for a 25 and 30 cells respectively.
	(d) and (h) ``shear'' 45$^{\circ}$ incident wave with respect to the slab's interface for a 25 and 30 cells respectively.}
\label{fig:30_60_cell}
\end{figure}

Indeed, by comparison of Fig.~\ref{fig:30_60_cell} with Fig.~(\ref{fig:20_cell}), it is possible to infer that the fact of increasing the slab's thickness mainly acts on the number and amplitude of oscillations that occur in the reflection coefficient for frequencies lower and higher than the band-gap.
Moreover, it can be seen from these figures that the performances of the relaxed micromorphic model is improved when increasing the number of unit cells constituting the metamaterial slab embedded in the Cauchy material \textit{CM1}.
Nevertheless, some mismatch can still be observed for frequencies higher than the band-gap, also for relatively high number of unit cells.
This high-frequency mismatch is related to the fact that the wavelength of the incident wave exceeds the threshold value $\ell_t$ as discussed before in Subsection~\ref{subsec:reflect_elastc_properties_outer}.

\begin{figure}[H]
	\centering
	\includegraphics[width=0.8\textwidth]{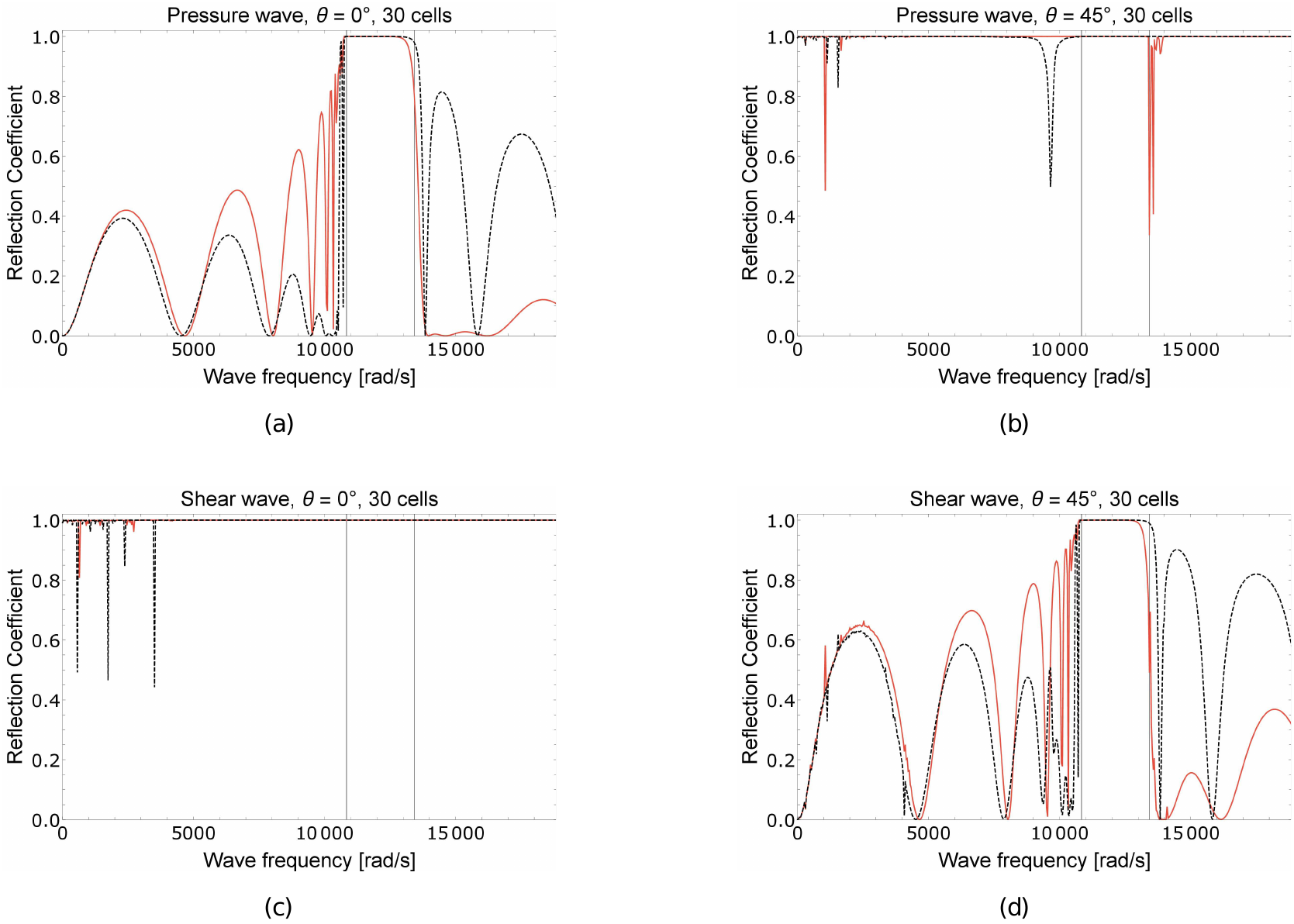}
	\caption{Comparison of the microstructure's (black line) and micromorphic (red line) reflection coefficient as a function of frequency for a 30 unit cells slab of \textit{MM1} and embedded in \textit{CM2}.
		(a) ``pressure'' normal incident wave with respect to the slab's interface.
		(b) ``pressure'' 45$^{\circ}$ incident wave with respect to the slab's interface.
		(c) ``shear'' incident wave normal to the slab's interface.
		(d) ``shear'' 45$^{\circ}$ incident wave with respect to the slab's interface.}
	\label{fig:30_cell_P_-1}
\end{figure}

To improve the higher-frequency micromorphic description of the structure in this case, a substantial constitutive extension of the relaxed micromorphic model is needed so as to account for higher frequency modes that presumably play an important role in this frequency range.
Similar arguments are valid for the slab embedded in the Cauchy material \textit{CM2}, as shown by Fig.~\ref{fig:30_cell_P_-1} and Fig.~\ref{fig:20_cell_P_-1}.
As for the slab embedded in the Cauchy material \textit{CM3}, we already observed in Fig.~\ref{fig:20_cell_SW} that its reflective behaviour is better caught by the relaxed micromorphic model than in the previous case, even at higher frequency.

\begin{figure}[H]
	\centering
	\includegraphics[width=0.8\textwidth]{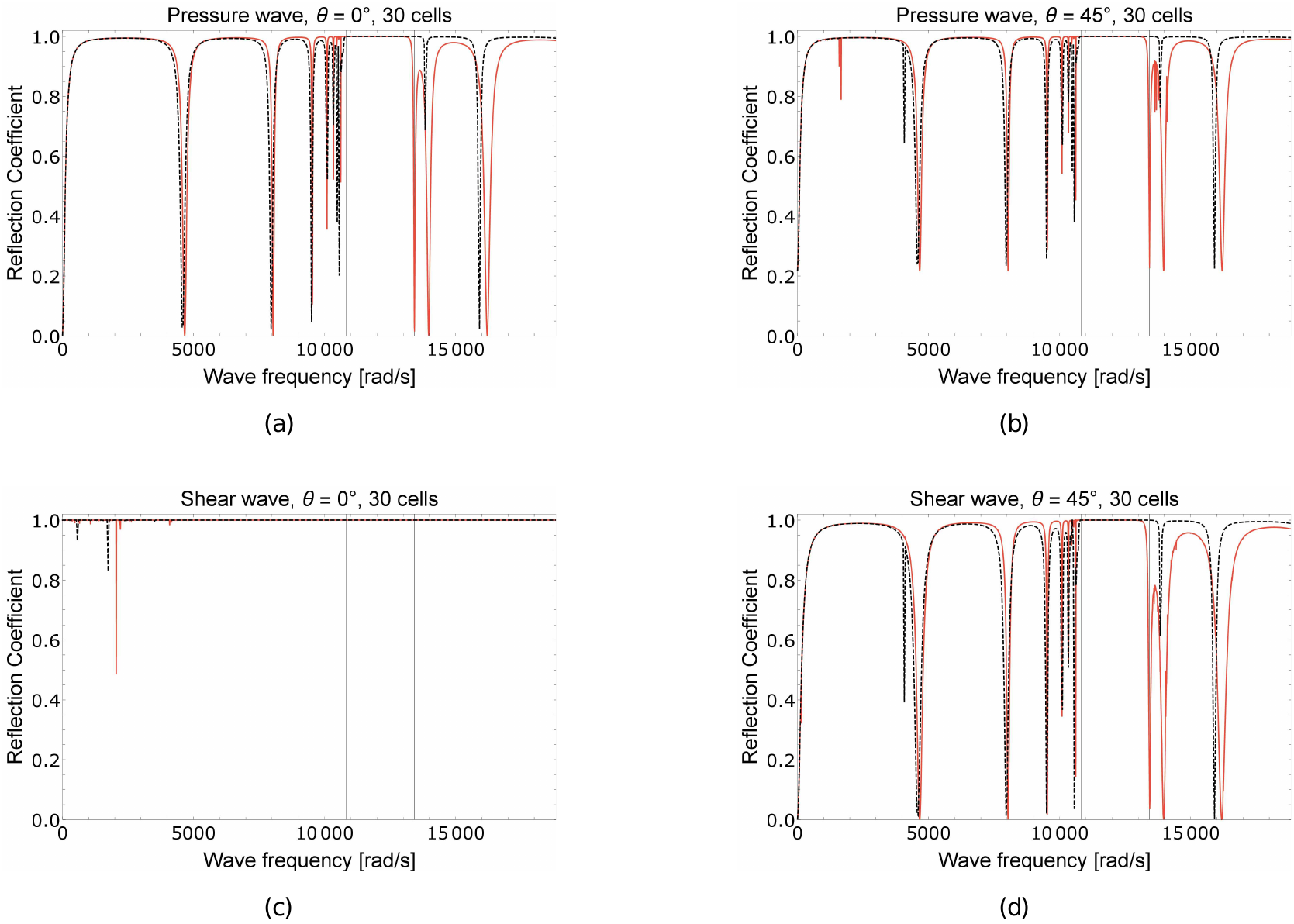}
	\caption{
	Comparison of the microstructure's (black line) and micromorphic (red line) reflection coefficient as a function of frequency for a 30 unit cells slab of \textit{MM1} and embedded in \textit{CM3}.
	(a) and (e) ``pressure'' normal incident wave with respect to the slab's interface for a 25 and 30 cells respectively.
	(b) and (f) ``pressure'' 45$^{\circ}$ incident wave with respect to the slab's interface for a 25 and 30 cells respectively.
	(c) and (g) ``shear'' incident wave normal to the slab's interface for a 25 and 30 cells respectively.
	(d) and (h) ``shear'' 45$^{\circ}$ incident wave with respect to the slab's interface for a 25 and 30 cells respectively.}
\label{fig:30_cell_SW}
\end{figure}

This means that, for this structure, the relaxed micromorphic model presented is sufficient for its correct description in the considered frequency range and that its generalization can be avoided in this case, also for relatively high frequencies.
Indeed, in this last case, the fact of increasing the number of unit cells does not significantly improve the description of the slab's refractive behaviour (see Fig.~\ref{fig:30_cell_SW}).
The slight differences between the reflection patterns obtained via the relaxed micromorphic model and those obtained via the full simulations (see Fig.~\ref{fig:30_cell_SW}) can hence be attributed uniquely to a constitutive enhancement of the relaxed micromorphic model to include extra degrees of freedom and higher modes.   

\subsection{Reflection coefficient of the tetragonal metamaterial slab as a function of the angle of incidence $\theta$ and of the wave-frequency}
Since the predictive capability of the relaxed micromorphic model here proposed has been assessed in the previous sections, in this subsection, we show the behaviour of the reflection coefficient as a function of the frequency of the incident wave and the angle of incidence for the structure of Fig.~\ref{fig:fig_tab_unit_cell}(a) with outer Cauchy material \textit{CM1}.

\begin{figure}[H]
	\centering
	\includegraphics[width=0.8\textwidth]{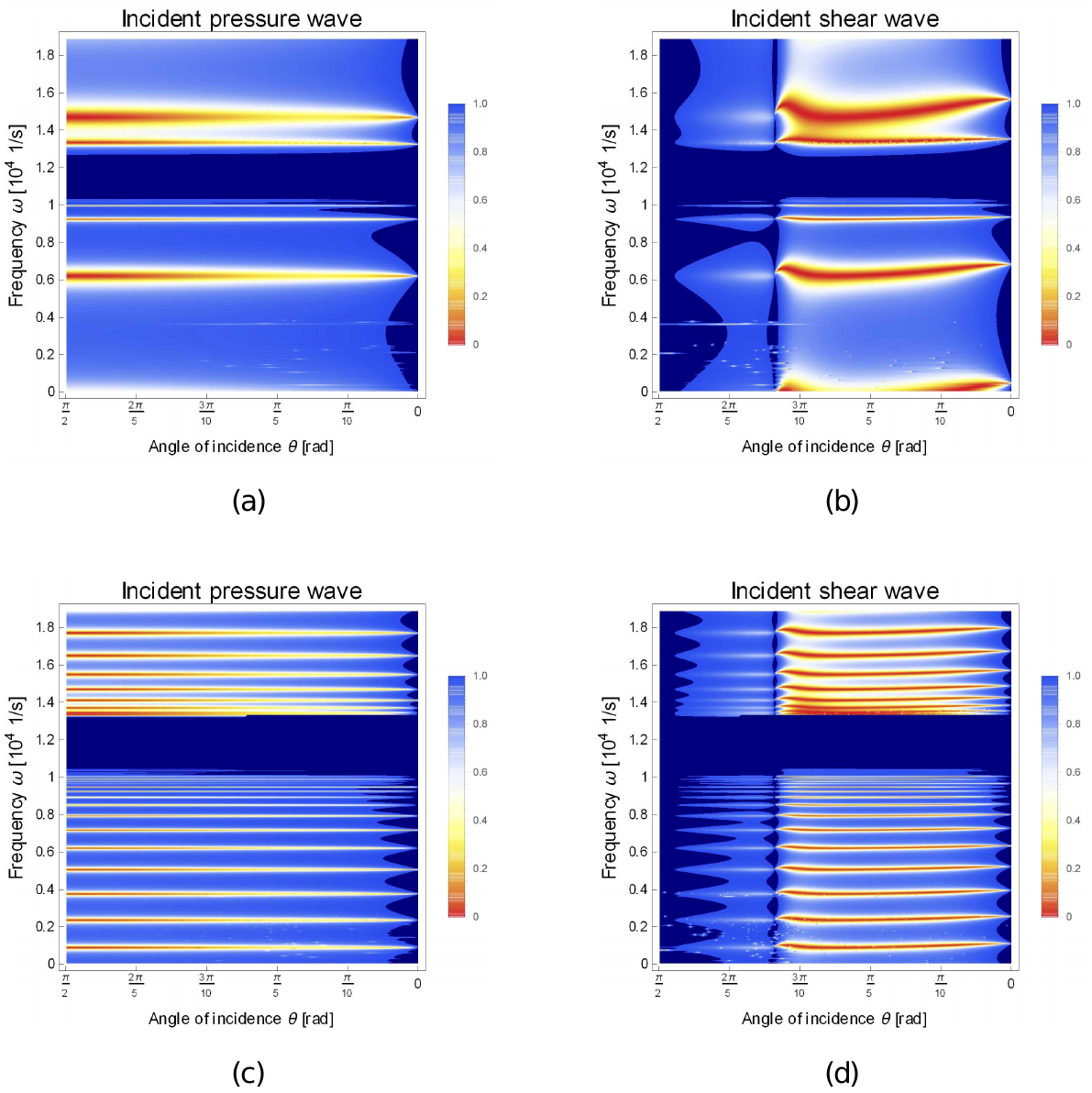}	
	\caption{
	Simulation of the reflection coefficient with the obtained relaxed micromorphic model for a 20 (a)-(b) and a 100 (c)-(d) unit cells thick slab made up of \textit{MM1} material and embedded in \textit{CM1} Cauchy as function of the angle of incidence and of the wave-frequency - (a) and (c): incident pressure wave; (b) and (d) incident shear wave.}
\label{fig:sweep_20_100_cell_ti}
\end{figure}

We start noticing that for the case of an incident ``pressure'' wave the structure's refractive behaviour is relatively unaffected by the value of the angle of incidence (see Fig.~\ref{fig:sweep_20_100_cell_ti}(a) and (c)).

In particular, when considering fewer unit cells in the metamaterial slab (Fig.~\ref{fig:sweep_20_100_cell_ti}(a)), we can observe almost total reflection occurring in a wide frequency range (extending outside the band-gap) and for all angles of incidence. Few frequencies can be identified around which total transmission occurs. When increasing the number of cells in the metamaterial slab, the frequencies around which total transmission occurs increase in number.
We can thus remark that the simple fact of considering a finite-size metamaterial with a different number of unit cells modifies the structure's behaviour in a significant way.

Things are even more interesting when considering ``shear'' incident waves, since the structure's behaviour starts being significantly affected by the angle of incidence of the travelling wave.

In particular, a critical angle exists (see Fig.~\ref{fig:sweep_20_100_cell_ti}(b) and (d)) such that all waves hitting the interface with an angle included between normal incidence and this critical value are almost completely reflected for any frequency (even outside the band-gap).
For angles beyond the critical value the structure's behaviour becomes similar to that observed for incident ``pressure'' waves.

\begin{figure}[H]
	\centering
	\includegraphics[width=0.8\textwidth]{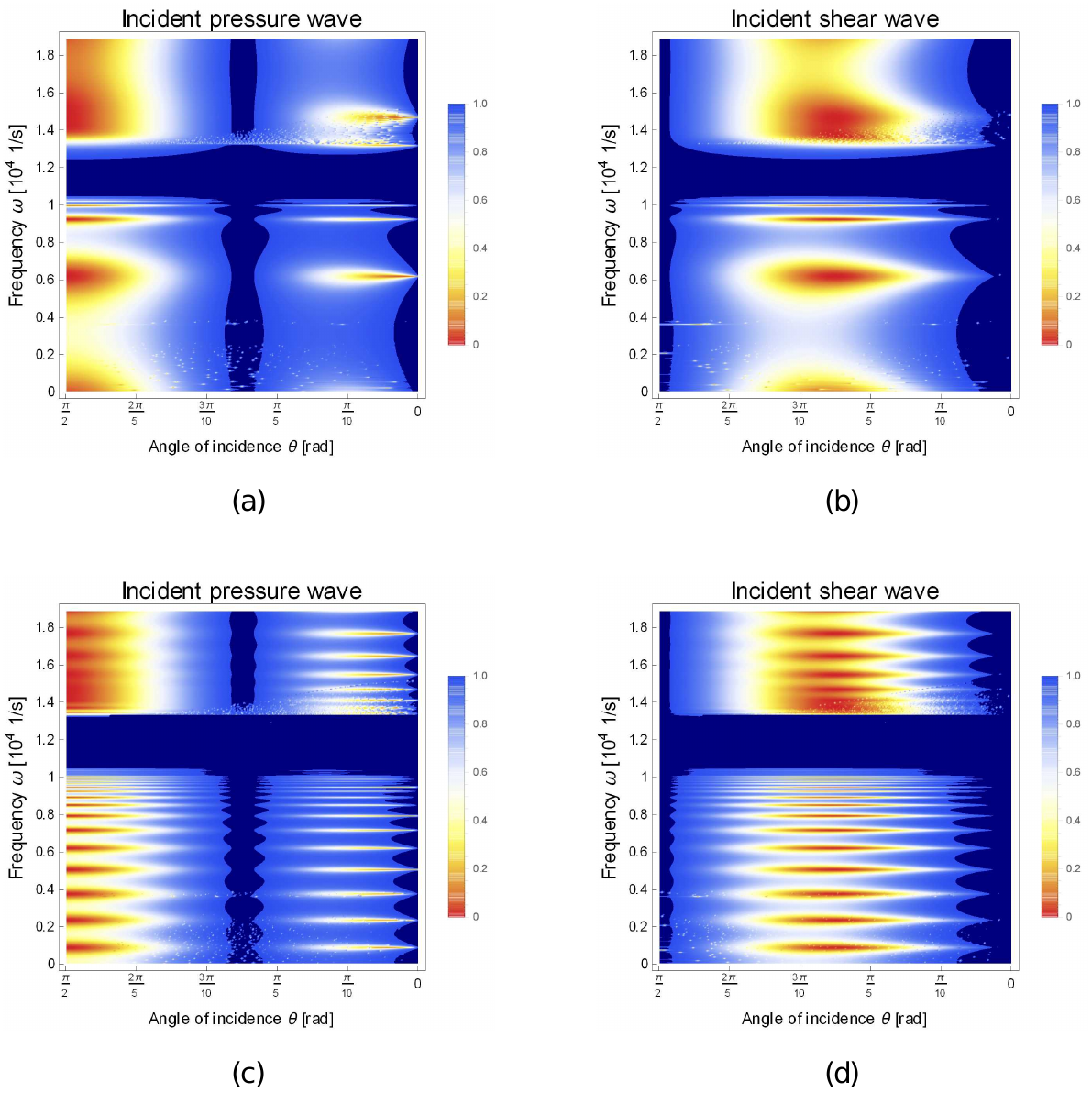}
	\caption{
	Simulation of the reflection coefficient with the obtained relaxed micromorphic model for a 20 (a)-(b) and a 100 (c)-(d) unit cells thick slab made up of \textit{MM1} material and embedded in \textit{CM2} Cauchy as function of the angle of incidence and of the wave-frequency - (a) and (c): incident pressure wave; (b) and (d) incident shear wave.}
\label{fig:sweep_20_100_cell_neg}
\end{figure}

We analyse the same meta-structure of Fig.~\ref{fig:fig_tab_unit_cell}(a) by now considering the material \textit{CM2} as ``outer'' Cauchy material. By direct comparison of Figs.~\ref{fig:sweep_20_100_cell_neg} with Figs.~\ref{fig:sweep_20_100_cell_ti}, it can be easily inferred that the meta-structure's behaviour is somehow reversed with respect to the previous structure.

First of all, we can identify a ``critical angle region'' for incident ``pressure'' waves instead than for ``shear'' ones. Moreover, we can remark that an almost total transmission occurs for angles smaller than this critical value instead than a total reflection in the previous case.

For ``shear'' incident waves the behaviour is still different because different zones are identified depending on the value of the incident angle.
In particular, two critical angles exist in this case such that total transmission occurs between these two critical values, while total reflection occurs otherwise.

In summary, we can see how the fact of simply changing the properties of the ``outer'' Cauchy material reverses the meta-structure's behaviour in specific frequency and angle-of-incidence ranges, both for ``pressure'' and ``shear'' incident waves.
\section{Conclusions}
In this paper we use the relaxed micromorphic model to characterize three different 2D tetragonal metamaterials that can be used for applications in acoustic control.
The reduced structure of the relaxed micromorphic model allows us to efficiently explore different meta-structural configurations in which a metamaterial's slab is embedded in a homogeneous Cauchy material.
As a result, we are able to show that the metamaterial's refractive behaviour can be drastically changed by simply acting on the stiffness of the homogeneous material.
In this way, the same structure can be adapted so as to act as a total screen or a total absorber in specific frequency and angle-of-incidence ranges.
The results presented so far, clearly show that the study of the mechanical behaviour of metamaterials cannot be disjoined by the study of their interactions with other materials, if one wants to enable the realistic conception of new engineering meta-structures.
By presenting our results, we also outline that, as any model, also enriched models have limitations that have to be identified "a priori" to avoid their inappropriate use.
In particular, we underline that the model's performances may depend on three characteristic lengths that are related to i) the unit cell's size, ii) the metamaterial's slab thickness and iii) the wavelength of the travelling wave. Depending on the relative proportions of these three characteristic lengths the relaxed micromorphic model will be more or less efficient in the description of the meta-structure's behaviour over an extended frequency range. Indeed, while the relaxed micromorphic model will always be predictive of this behaviour in the long wave limit, more or less marked differences may emerge for higher frequencies and lower wavelengths.
This calls for the formulation of a new enriched model, including extra degrees of freedom and suitable dynamic internal lengths, so as to extend its efficiency to higher-frequency/smaller-wavelength regions for a wide panel of external excitations.

\let\oldbibliography\thebibliography
\renewcommand{\thebibliography}[1]{%
\oldbibliography{#1}%
\setlength{\itemsep}{1.7pt}%
}
\begin{scriptsize}
	\bibliographystyle{plain}
	\bibliography{library}
\end{scriptsize}
\end{document}